\documentclass{aastex}
\usepackage{emulateapj5}

\slugcomment{accepted to ApJ} 

\shorttitle{X-ray Spectroscopy of Her X-1: CNO Abundances}
\shortauthors{Jimenez-Garate et al.}

\begin{document}


\title{High-Resolution X-ray Spectroscopy of Hercules X-1 with the {\it XMM-Newton} RGS: \
CNO Element Abundance Measurements and Density Diagnostics of a Photoionized Plasma}


\author{M. A. Jimenez-Garate,\altaffilmark{1,2} C. J. Hailey,\altaffilmark{2} J. W. den Herder,\altaffilmark{3} S. Zane,\altaffilmark{4} and G. Ramsay\altaffilmark{4}}

\altaffiltext{1}{Center for Space Research, Massachusetts Institute of
Technology, 70 Vassar St, NE80-6091, Cambridge, MA, 02139; mario@space.mit.edu.}
\altaffiltext{2}{Columbia Astrophysics Laboratory, New York, NY, 10027;
chuckh@astro.columbia.edu.}
\altaffiltext{3}{SRON, the National Institute for Space Research, The
Netherlands; J.W.A.den.Herder@sron.nl.}
\altaffiltext{4}{Mullard Space Science Laboratory, University College of London, UK;
sz@mssl.ucl.ac.uk, gtbr@mssl.ucl.ac.uk.}



\begin{abstract}
We analyze the high-resolution X-ray spectrum of Hercules
X-1, an intermediate-mass X-ray binary, which was observed with the
{\it XMM-Newton} Reflection Grating Spectrometer. We measure the
elemental abundance ratios by use of spectral models, and we detect
material processed through the CNO-cycle.  The CNO abundances,
and in particular the ratio N/O~$> 4.0$ times solar,
provide stringent constraints on the evolution of the binary
system.  The low and short-on flux states of Her X-1 exhibit narrow
line emission from \ion{C}{6},
\ion{N}{6}, \ion{N}{7}, \ion{O}{7}, \ion{O}{8}, \ion{Ne}{9}, and
\ion{Ne}{10} ions.
The spectra show signatures of photoionization.
We measure the electron temperature, quantify
photoexcitation in the He$\alpha$ lines, and set limits on the
location and density of the gas. The recombination lines may originate
in the accretion disk atmosphere and corona, or on the X-ray
illuminated face of the mass donor (HZ~Her). 
The spectral variation over the course of the 35~d
period provides additional evidence for the precession of the disk.
During the main-on state, the narrow line emission is absent, but we
detect excesses of emission at $\sim$10--15~\AA, and also near the \ion{O}{7}
intercombination line wavelength. 

\end{abstract}


\keywords{X-rays: binaries --- line: formation --- line: identification --- pulsars: individual (Her X-1) --- 
accretion, accretion disks --- binaries: eclipsing}


\section{Introduction}
Hercules X-1 is a bright intermediate-mass X-ray binary which has been
observed extensively after its discovery by
\citet[]{first_disc}. The system contains an
X-ray pulsar with $P_{\rm pulse} =1.24$~s period \cite[]{discovery_pulsar,batse}.
A synchrotron resonance feature yields a magnetic field of $B = 3.5
\times 10^{12}$~G \cite[]{beppo_cyc,truemper_cyc}. 
Some models yield a lower magnetic field \cite[]{lowfield}.
Optical light-curves \cite[]{optical_curves}
and X-ray eclipses \cite[]{discovery_pulsar} yield
a $P_{\rm orb} = 1.7$~day orbital period.
The $1.5 \pm 0.3~M_{\sun}$ neutron star has a $2.3 \pm 0.3~M_{\odot}$ companion, HZ~Her, 
which changes from A to B spectral-type over the orbital period, due to the strong X-ray
illumination on its surface \cite[]{reynolds}.
Pre-eclipse and anomalous dips in the X-ray flux are observed 
\cite[]{dipsfound,xteobs} due to the interaction of an accretion
stream with the accretion disk.  
The unabsorbed luminosity of Her X-1 is $L=3.8 \times
10^{37}$~ergs$^{-1}$, using a distance of $D = 6.6 \pm 0.4$~kpc 
\cite[]{reynolds}.

Hercules X-1 exhibits an unusual long-term X-ray flux modulation with
$P_{\Psi} = 35$~d period \cite[]{dipsfound}. Changes associated with
this period have also been observed in the optical light curves
\cite[]{optical_curves}, X-ray pulse shapes \cite[]{gingapulse},
X-ray dips \cite[]{dipsfound,xteobs}, and X-ray spectra \cite[]{ramsay}.
The 35~d X-ray light-curve is asymmetric and contains two maxima:
a state of $\sim 8$~d duration reaching the peak flux $F_{\rm max}$
named the {\it main-on},
and a secondary high state of $\sim 4$~d duration reaching $\sim 1/3~F_{\rm max}$ 
named the {\it short-on}. A low-flux state with $\sim 1/20~F_{\rm max}$ 
ensues at other epochs. The period $P_{\Psi}$ varies from cycle to cycle, and it
has been observed to be 19.5, 20, 20.5, and 21 times $P_{\rm orb}$ 
in the course of $\gtrsim 5$~yr of continuous monitoring 
with the All-Sky Monitor onboard the {\it Rossi X-ray Timing
Explorer} \cite[]{xteobs}.
The $\Psi = 0$ phase is defined as the time the main-on state begins.
Observations indicate that $\Psi =0$ coincides only with orbital
phases $\phi = 0.23$ or 0.68 \cite[]{xteobs}.

The 35~d cycle has been associated with a tilted accretion disk
that precesses by some unknown mechanism 
\cite[]{precessing_disk_hyp,twisted_disk_flaps,disk_model}.
The X-ray light-curve, the variations in the pulse profiles, and
the variability of the dips with $\Psi$-phase are fit by a geometric model
of a precessing, warped accretion disk, together with the beam
of a pulsar \cite[]{disk_model}. 
Among the parameters obtained in this fit are
an $85^{\circ}$ inclination with respect to the line of sight, 
a $20^{\circ}$ precession opening angle for the outermost disk,
and an $11^{\circ}$ precession angle for the innermost disk. 
The latter precession angles are dependent on assumed input parameters
such as the disk thickness. 

The ultraviolet (UV) spectrum exhibits line emission from species such as
\ion{C}{4}, \ion{N}{5}, and \ion{O}{5}, which originate in two separate
components that produce superimposed broad and narrow lines.  The intensity of the
narrow lines has the same orbital variation as the UV continuum, which is
thought to originate on the illuminated face of HZ~Her.
The broad line component follows roughly the
velocities expected for the accretion disk \cite[]{hut_uvlines}. 
Observations during eclipse ingress and egress
suggest that the broad line region originates in a prograde
disk of $\sim 10^{11}$ cm radius \cite[]{uv_disk_lines}.
The UV lines (i.e. \ion{N}{5}) have a weak P Cygni profile component
which indicates the presence of a wind outflow \cite[]{uv_lines_wind_model}.

The Her X-1 broadband X-ray emission can be described as a blackbody component
with temperature $kT \sim 90$~eV,  plus a power-law
component with a 24~keV exponential cutoff, and a
42~keV cyclotron feature \cite[]{beppo_cyc}. 
An Fe K fluorescence line evolves with $\Psi$-phase. 
The {\it XMM-Newton} EPIC data, presented by \cite{ramsay},
show a 6.4~keV Fe K line which is practically unresolved during the low and short-on states, 
and a broad line at 6.5~keV with $330 \pm 20$~eV FWHM during the main-on.

In this article, we present the high-resolution spectrum of Her X-1
in the 5 to 38 \AA \ band, which we obtained with the \it XMM-Newton \rm
Reflection Grating Spectrometer (RGS). The high-resolution X-ray
spectrum evolves dramatically with 35~d phase.  We analyze the
spectral data from three distinct epochs (section \ref{sec:obs}).
The spectrum during the low and short-on states is a faint power-law
continuum plus many narrow emission line features from a photoionized plasma (section
\ref{sec:lowshort}).  In contrast, the main-on state has a bright
continuum,
neutral absorption features,
excess of emission near \ion{O}{7} He$\alpha$,
and a continuum which cannot be fit well with blackbody and/or
power-law models (section \ref{sec:mainon}).
The observed emission lines arise from the recombination of electrons
with ions and from the subsequent radiative cascades. The gas is
predominantly heated by photons due to the large luminosity and small
volume of the system.  

We analyze and model the high-resolution spectra of Her X-1 to
obtain the elemental abundance ratios and measure the state variables of the gas, 
and we make preliminary identifications of the newly detected
emission region(s).  We apply spectral diagnostics on the plasma to constrain its location
(section \ref{sub:helike}), its density (section \ref{sub:param}), and to
measure the electron temperature (section \ref{sub:rrc}). We introduce
a method to extract the elemental abundance ratios from the low and
short-on spectra. The method involves extracting the emission measure
distribution from model fits of the recombination line fluxes (section
\ref{sec:em}). We discuss whether the narrow line emission originates
in a disk atmosphere and corona, on the illuminated
companion, or in both (section \ref{sub:em}).  We describe the
implications of the detection of CNO-processed material for the
evolution of the X-ray binary (section \ref{sub:evol}). We outline a
physical scenario which produces broad emission lines, which is
associated with the magnetopause and the inner disk in X-ray pulsars
(section \ref{sub:magn}). In that context, we speculate on the origin
of the complex continuum observed during the main-on (section
\ref{sub:bump}). Our conclusions are in section \ref{sec:concl}.

\section{Observations}
\label{sec:obs}

\begin{figure}
\plotone{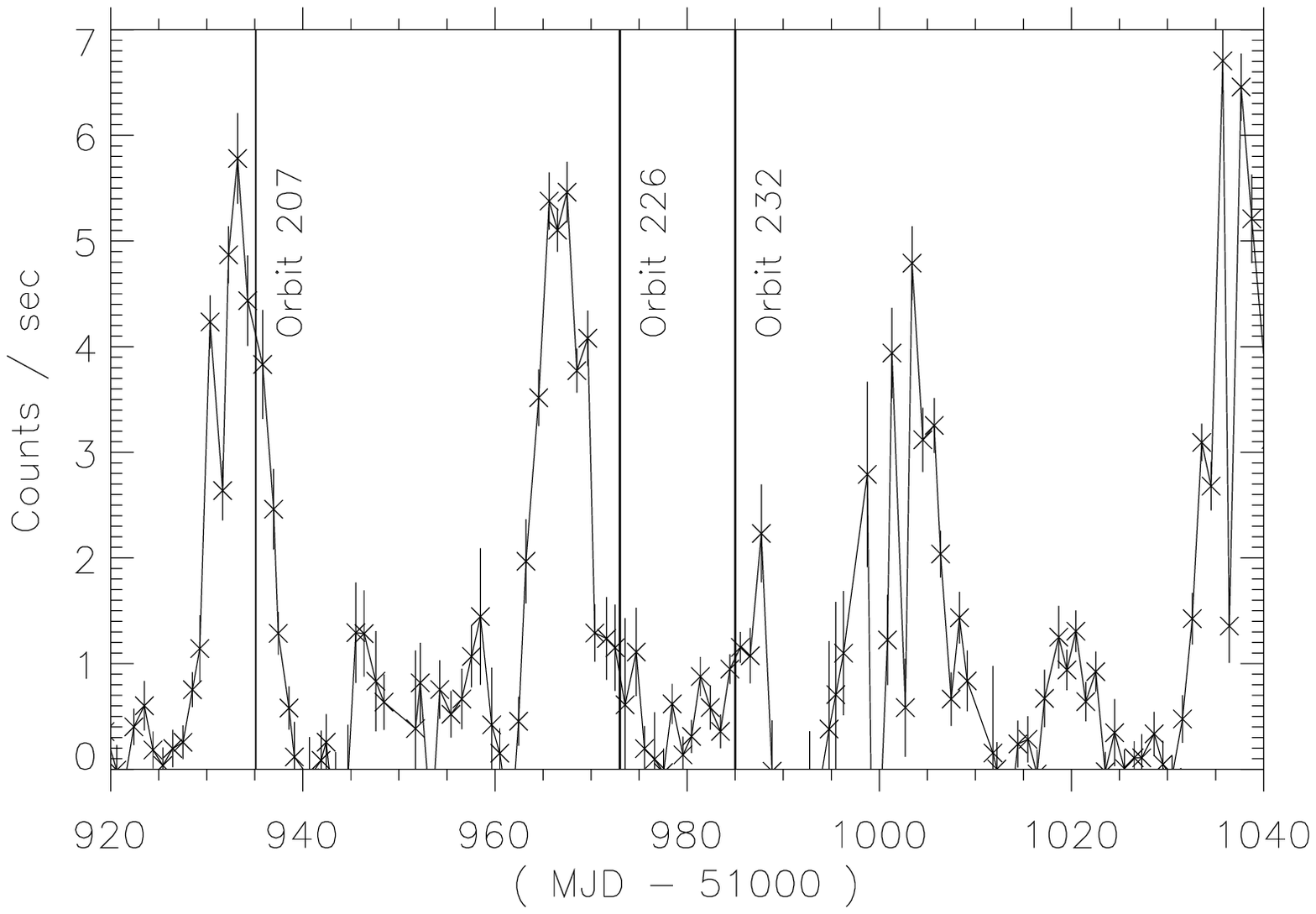}
\caption{Light curve obtained with the \it RXTE \rm 
all-sky monitor.  \label{fig:asm}}
\end{figure}

Three observations with \it XMM-Newton \rm RGS during the main-on,
short-on and low states, reveal X-ray line emission that varies with
$\Psi$-phase. The \it RXTE \rm  All-Sky Monitor ($ASM$) public data are
used to find the epoch of main-on start, or $\Psi=0$ (see the light
curve in Fig. \ref{fig:asm}).  The values of the $\phi$- and
$\Psi$-phases during the \it XMM-Newton \rm observations are shown in
Table \ref{tab:obs}. 

We describe the processing of each RGS spectrum.
All the data are processed with the most recent version of the
science analysis software ($SAS$ v. 5.3).  Response matrices are
extracted using 4000 energy bins in the full RGS energy range.
Hot pixels which are flagged by the
pipeline processing are excluded from the analysis.  Continuum and line fits are
performed with the XSPEC v.11.1 analysis software \cite[]{xspec}.
The total RGS background count-rate is 0.09~c~s$^{-1}$ in the low state,
0.08~c~s$^{-1}$ in the short-on state, and $\le 0.28$~c~s$^{-1}$ 
in the main-on state. The background is roughly independent of
wavelength, at a level of $\sim 2.7 \times
10^{-3}$~c~s$^{-1}$~\AA$^{-1}$ in the low and
short-on states. 
The background level rises by a factor of 2 to 4 for
wavelengths $\lambda < 7$~\AA.
The background level in the main-on is very small ($\le
0.015$~c~s$^{-1}$ \AA$^{-1}$) compared to scattering in the
cross-dispersion direction from first order photons.
The background-subtracted
event lists from the $SAS$ pipeline are used throughout. The extraction regions
from the RGS data are inspected and left unchanged from the $SAS$ default.
During main-on, RGS~2 is set
to high timing resolution mode, as part of the performance verification
of the instrument, while RGS~1 is set to spectroscopy mode. 
The RGS guaranteed time observations were performed during the low and
short-on states, with both RGS in spectroscopy mode.
We present the spectroscopy mode data.  
Since CCD chip 7 in RGS~1 and CCD chip 4 in RGS~2 are not operational, the
wavelength ranges of 10.3 to 13.6 \AA \ and 19.9 to 24.9 \AA \ are excluded from
each RGS dataset, respectively. For the main-on data,
we have enough counts to use the RGS~1 second order, which
fills the wavelength gap in the first order.

\section{Spectroscopy of the low and short-on states}
\label{sec:lowshort}

\begin{figure}
\epsscale{0.6}
\plotone{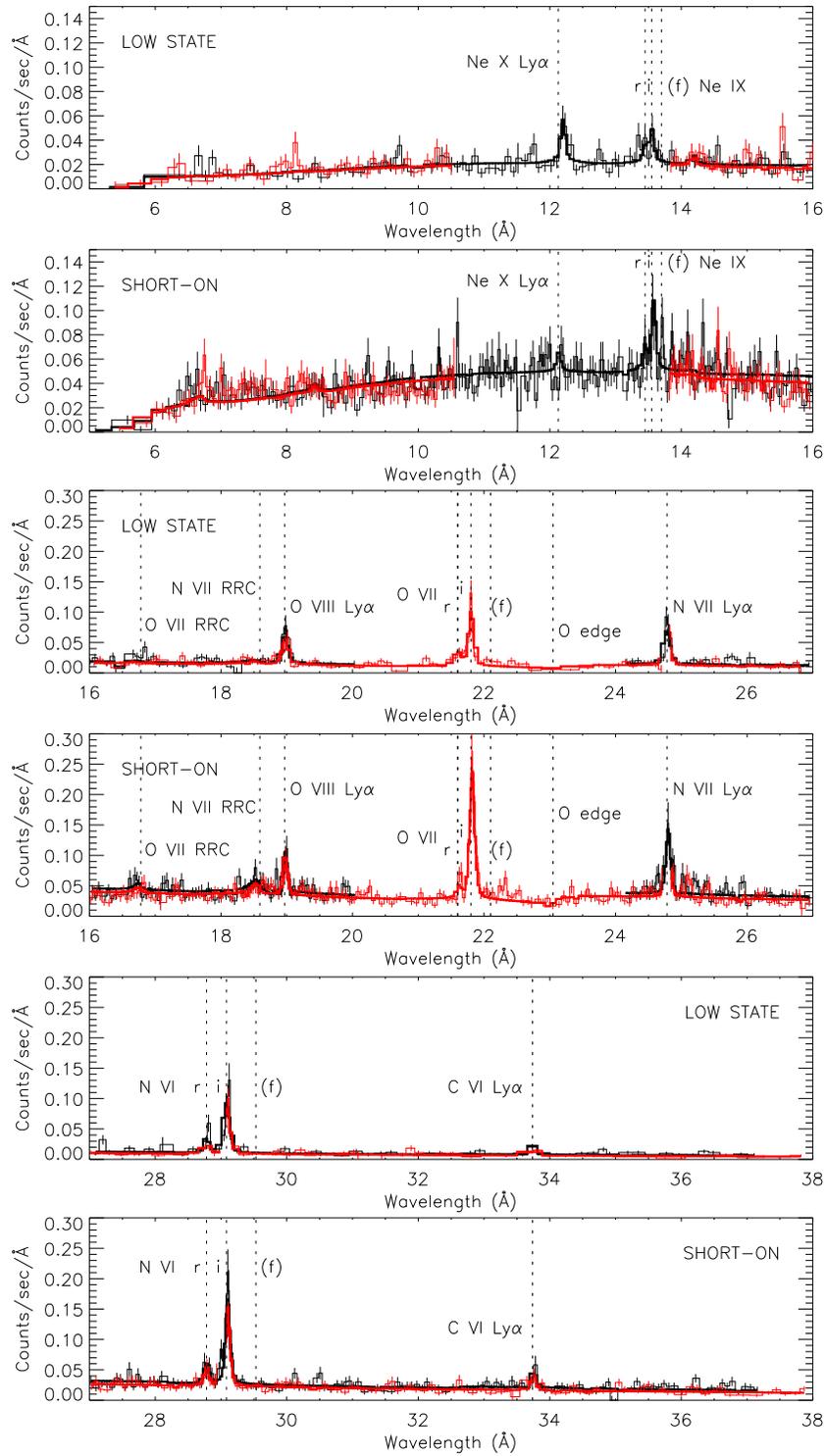}
\caption{High-resolution spectra obtained with RGS~1 (red) and RGS~2 (black),
and phenomenological fits (thick-lines).
The low state (13~ks) and short-on state (12~ks) observations are shown on adjacent panels.
The spectra have been rebinned adaptively. \label{fig:data_mod_lowshort}}
\end{figure}

The low-state RGS spectrum (shown in Fig. \ref{fig:data_mod_lowshort})
contains line emission from \ion{Ne}{9}, \ion{O}{8}, \ion{O}{7},
\ion{N}{7}, \ion{N}{6} and \ion{C}{6} ions.
Within the resolution, no velocity broadening has been observed.
The continuum shape is described by a power-law,
although the goodness-of-fit is likely degraded by
a collection of weak and unresolved spectral lines.
There is a neutral O edge, which is mostly instrumental. The line and
continuum measurements are shown in Tables \ref{tab:lines} and 
\ref{tab:cont}, respectively. 
No evidence for absorption local to the source is observed on these states, implying
that the line emission region is optically thin to both lines and
continuum.

\begin{figure}
\epsscale{1}
\plotone{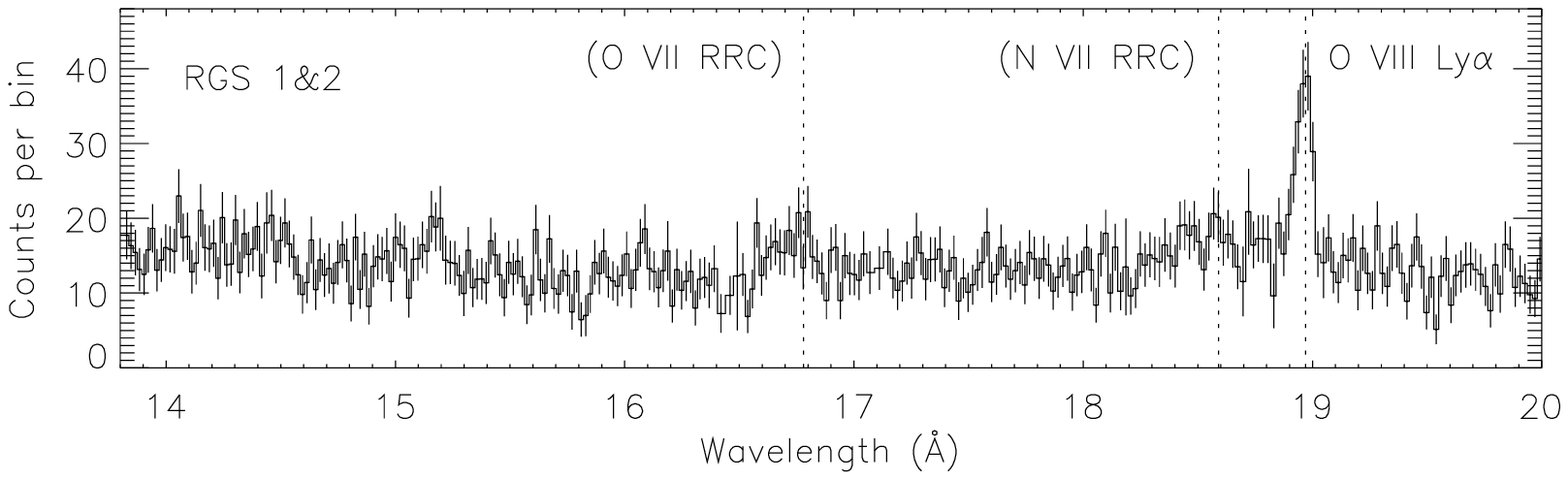}
\plotone{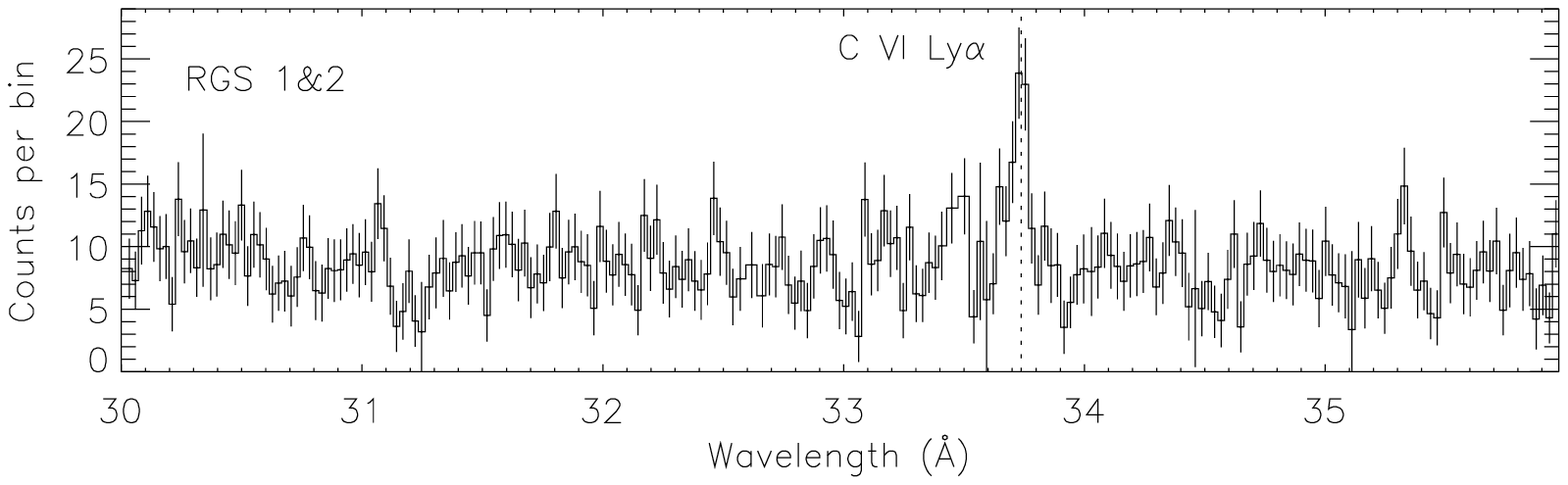}
\caption{Summed
RGS spectra of both low-state and short-on data, which 
highlight the observed RRC of \ion{O}{7} and \ion{N}{7},
and \ion{C}{6} Ly$\alpha$.  
The bin size is 21 and 27 m\AA \ for the top and bottom frame,
respectively.  The line spread function is different for each RGS and
is not corrected on these spectra.
\label{fig:added_spec}}
\end{figure}

The total RGS count-rate in the short-on is two times larger than during the
low-state (also in Fig. \ref{fig:data_mod_lowshort}).
There is variability within each observation as well, but we
leave that out of our analysis.  
The line luminosities are usually larger in
the short-on than in the low state (Table \ref{tab:lines}).  
The better statistics during the
short-on allows us to detect the \ion{O}{7} and
\ion{N}{7} radiative recombination continua (RRC).
On Figure \ref{fig:added_spec}, we show the added spectrum of the low
and short-on states, which have similar features, to clarify the
statistical significance of the detection 
of \ion{Ne}{9} He$\alpha$, \ion{C}{6} Ly$\alpha$, the \ion{O}{7} RRC,
and the \ion{N}{7} RRC (the line profiles in this figure are not adequate for 
analysis because we have added spectral data with
different spectral resolution).  The EPIC 0.3--10~keV count rates, which are
398, 54 and 21~c~s$^{-1}$ in the main-on, short-on, and low states, respectively
\cite[]{ramsay}, indicate fluxes which are consistent with
the long term \it XTE \rm $ASM$ light curve
\cite[]{xteobs}, with our main-on observation near peak, and our
short-on observation at half-peak.

Our phenomenological model fit to the spectrum consists of a power-law continuum, 
emission lines, absorption edges, and RRC.  To perform the continuum
fits, and to use the $\chi^2$ statistic, we bin the spectrum with a
minimum of 20 counts per bin.
The spectra shown in Figures \ref{fig:data_mod_lowshort},
\ref{fig:added_spec}, and \ref{fig:high} are rebinned as indicated 
on each caption. The spectral lines are fit at full resolution with Gaussian functions
by independently varying their normalization, width, and centroid. The
line identifications are shown in Table \ref{tab:lines}.
The continuum is re-fit for accuracy after the line fits are added.
The ISIS analysis software \cite[]{isis} is used to check the line fits
from XSPEC. With ISIS, the continuum level is measured  
in two regions adjacent to the lines. The line equivalent widths ($EW$)
calculated from that continuum level are shown in the last two columns
of Table \ref{tab:lines}.  
The line fluxes from XSPEC and the equivalent widths
measured with ISIS agreed to $< 10$ \%, although the
difference is $\sim 25$ \% for
\ion{N}{7} He$\alpha$, \ion{N}{6} He$\alpha$, and \ion{O}{8} Ly$\alpha$.
The difference is due to some hot pixels which are removed only with XSPEC,
and to the uncertainty of the continuum flux level.

\subsection{Helium-like ion emission line diagnostics}
\label{sub:helike}

Electron density and temperature measurements in the $10^8 < n_e < 10^{18}$~cm$^{-3}$
and $10^6 < T < 10^7$ K ranges can be obtained with
helium-like ion emission line ratios \cite[]{gabriel,porquet}.
The dominant ionization mechanism can be determined from the same lines
\cite[]{liedahl99}. Low-Mass X-ray binaries (LMXB), and other
X-ray binaries with accretion disks, should exhibit 
densities high enough to render these diagnostics useful.

\begin{figure}
\epsscale{0.3}
\plotone{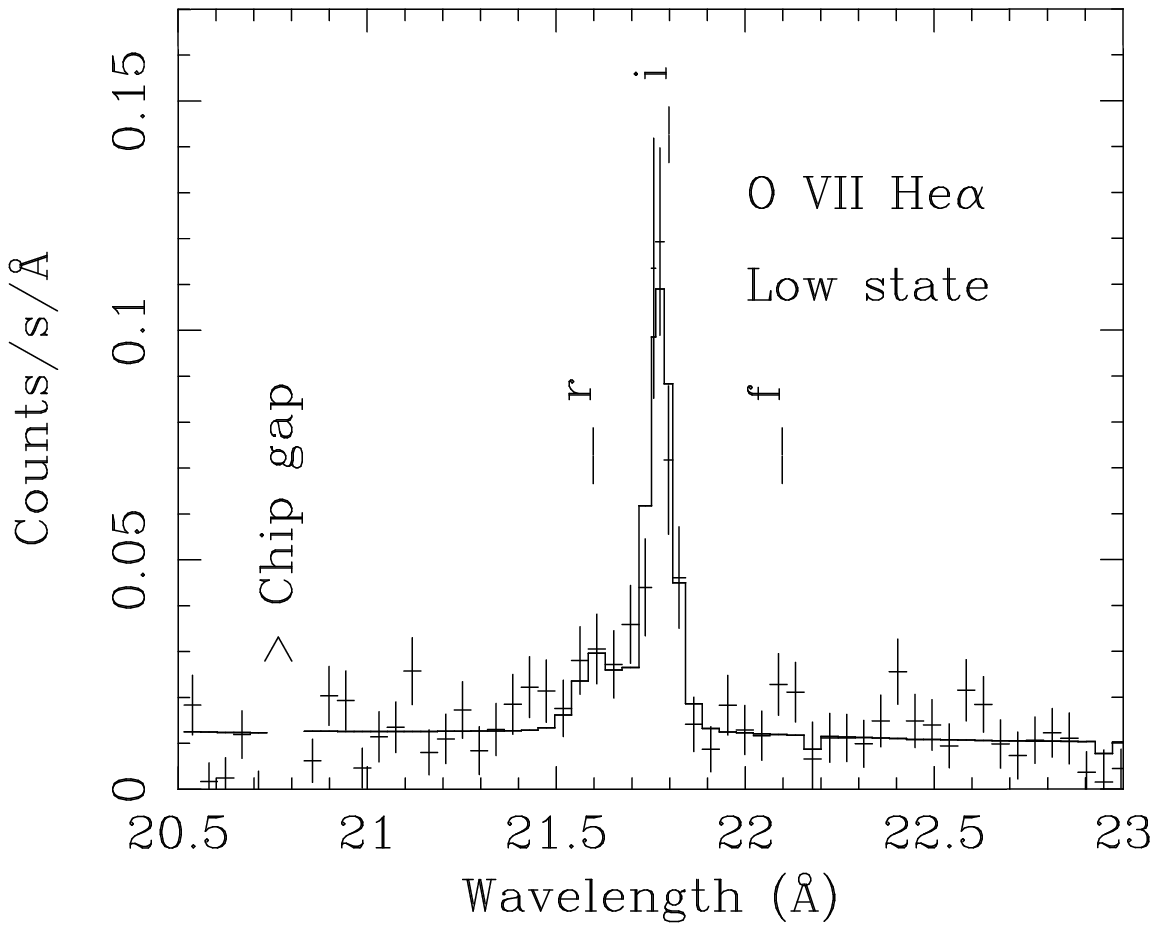}
\plotone{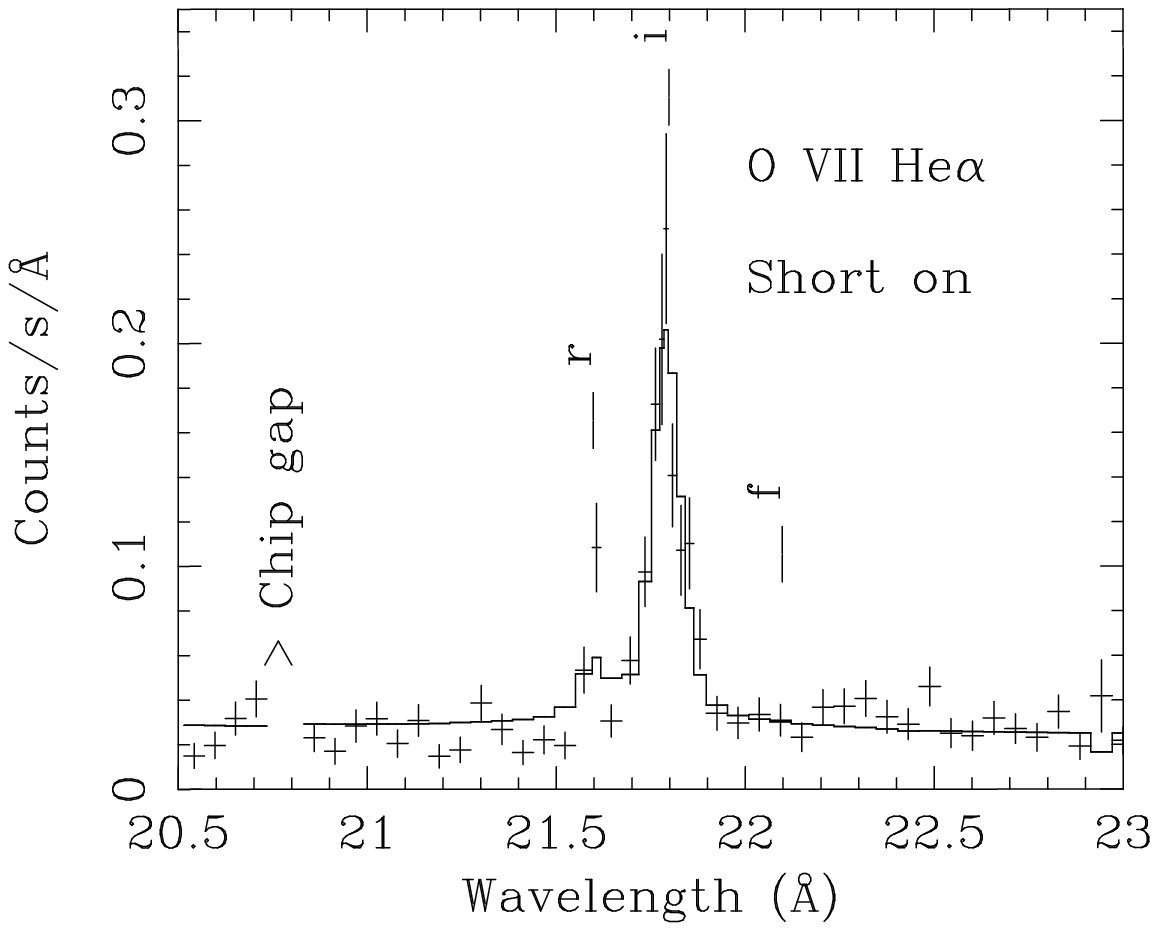}
\plotone{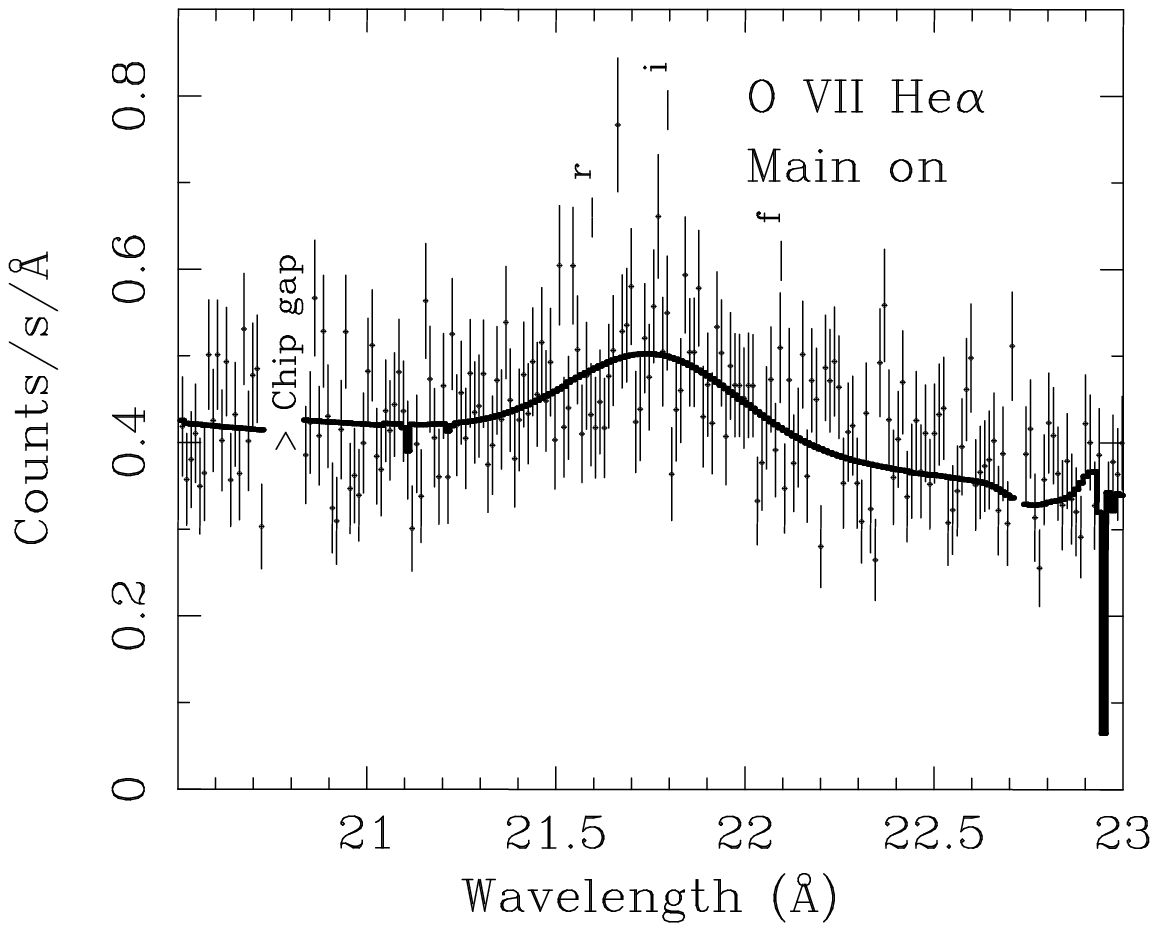}
\caption{The evolution of the \ion{O}{7} emission line 
complex with 35~d phase. The intercombination ($i$) and resonance ($r$)
lines are present in the low and short-on states. A complex, broad structure during 
the main-on is plotted at full resolution. 
RGS~1 data (symbols) and fits (solid lines) are shown.
\label{fig:ovii}}
\end{figure}

\begin{figure}
\epsscale{0.4}
\plotone{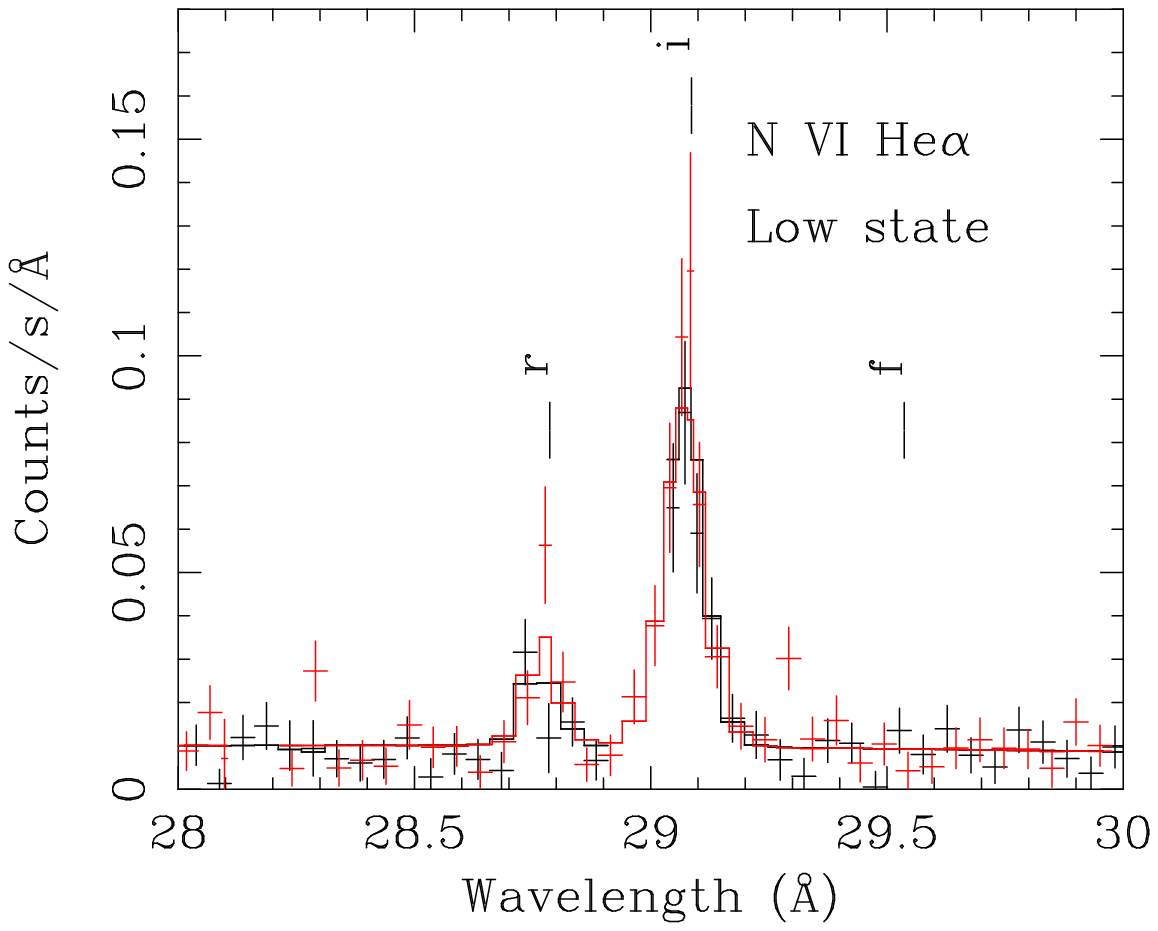}
\plotone{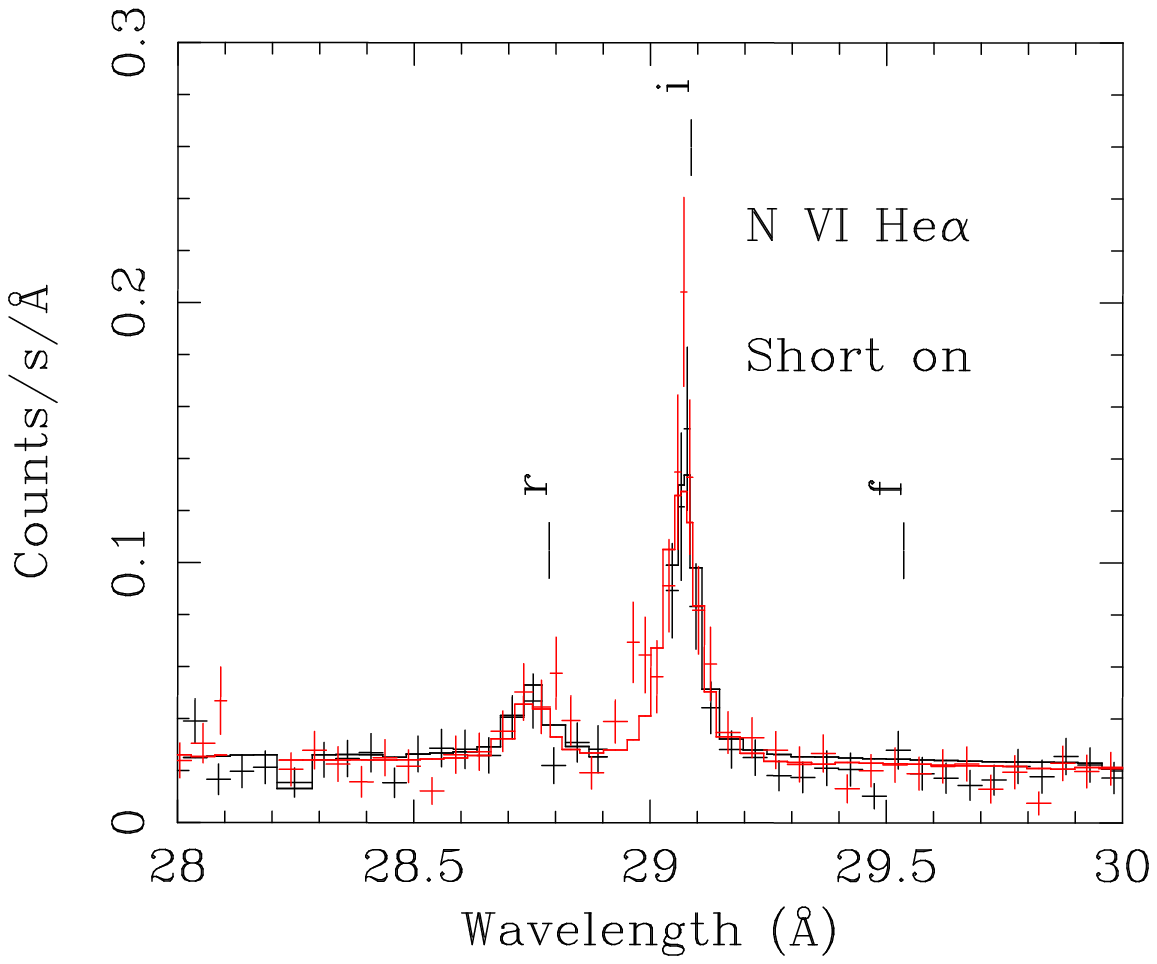}
\caption{The evolution of the \ion{N}{6} emission line 
complex with 35~d phase. The intercombination ($i$) and resonance ($r$)
lines are present in both low and short-on states. 
RGS~1 data (black symbols), RGS~2 data (red symbols), and the respective 
fits (solid lines), are shown.
\label{fig:nvi}}
\end{figure}

In Her X-1, the helium-like ion line complexes show a prominent intercombination 
line blend ($i$ denotes the two transitions $2^3P_{1,2} \to \ 1^1S_0$)
and a weaker resonance line ($r$ denotes $2^1P_1 \to 1^1S_0$),
as shown in Figures \ref{fig:ovii} and \ref{fig:nvi}. 
The forbidden line ($f$ denotes $2^3S_1 \to 1^1S_0$) is not detected for
any of the helium-like ions \ion{Ne}{9}, \ion{O}{7}, and \ion{N}{6}.
The notation $n^{2S+1}L_{J}$ identifies each two-electron atomic state,
with quantum numbers $n$, $L$, $S$, and $J$.

The line ratio $G=(f+i)/r$ is consistent with $\simeq 4$
for all cases shown in Table \ref{tab:helike}, which
indicates that photoionization is the dominant heating mechanism. For collisionally
ionized gases, $G \lesssim 1$, decreasing for larger temperatures
\cite[]{liedahl99,porquet}. The $G$ ratio indicates the electron
temperature is $T < 2 \times 10^6$~K, or $kT < 170$~eV \cite[]{porquet}.

We set upper limits for the $R=f/i$ line ratio
(see Tables \ref{tab:lines} and \ref{tab:helike}).
The $R \sim 0$ ratio indicates that 1) the density is above the threshold calculated
by \citet{porquet} and \citet{bauti}, thereby exciting the $2^3S_1$ level to the
$2^3P_{1,2}$ levels by electron impact, and/or 2) the $2^3S_1$ level
is photo-excited to the same $2^3P_{1,2}$ levels,
as described by \citet{mewe}, and later identified in
an O-star wind by \citet{kahn}. The density limits implied by the measured $R$,
ignoring photoexcitation, would be
$n_e > 4 \times 10^{11}$~cm$^{-3}$ for \ion{N}{6},
$n_e > 2 \times 10^{12}$~cm$^{-3}$ for \ion{O}{7}, and
$n_e > 5 \times 10^{12}$~cm$^{-3}$ for \ion{Ne}{9} \cite[]{porquet}.
However, the photoexcitation rate is
\begin{equation}
w_{\rm f \to i} = \frac{\pi e^2}{ m_e c} F_{\nu_{f \to i}} f_{\rm osc},
\end{equation}
where $e$, $m_e$ are the electron charge and mass, $c$ is the speed of light,
$f_{\rm osc}$ is the oscillator strength, and
$F_{\nu_{f \to i}}$ is the flux (in photons~s$^{-1}$~cm$^{-2}$~Hz$^{-1}$)
at frequencies resonant with the $2^3S_1 \to 2^3P_{1,2}$ transitions, in
this case in the UV.  We use the $f_{\rm osc}$ calculated by \citet{cann}. 
The calculated rates (Table \ref{tab:helike}) imply that UV
photoexcitation can compete with electron impact excitation, such that
the $R$ ratio density diagnostics are not directly applicable in Her
X-1.

Instead, we use the $R$ ratio to set limits on the distance 
between the X-ray line region and the UV source, $d$.
Because the radiation field is diluted with increasing distance,
$w_{\rm f \to i} \propto F_{\nu_{f \to i}} \propto 1/d^2$.
Beyond a threshold distance $d_{\rm crit}$, which is set by $w_{\rm f \to i}=w_f$,
the photoexcitation rate $w_{\rm f \to i}$ is smaller than the
radiative decay rate $w_f$, so
the $f$ line is not affected by photoexcitation, and neither
is $R$. For $d < d_{\rm crit}$, photoexcitation can suppress the $f$
line and enhance the $i$ line, until the $R$ ratio approaches zero
when the UV radiation field is sufficiently strong.
To calculate $d_{\rm crit}$, we use the distance to Her X-1 of $D =
6.6$~kpc \cite[]{reynolds}, the UV flux $F_{\nu_{f \to i}}$ measured 
by \cite{fos} and \cite{ghrs}, and we assume a UV point source.
Our results are shown in Table \ref{tab:helike}, 
which yield $d_{\rm crit} \sim 10^{12}$ cm.  
\cite{kahn} used this sort of procedure for a collisionally ionized gas.
% revised
The value of $d_{\rm crit}$ becomes $\sim 1/2$ of that in Table
\ref{tab:helike} if we use the $F (\lambda = 1637~$\AA$) \sim 0.4 \times
10^{-13}$~erg~s$^{-1}$~cm$^{-2}$~\AA$^{-1}$ measured by
\citet{hut_uvlines} at $\phi = 0.66$. This flux is in rough agreement
with observations by \citet{uv_disk_lines} at $\phi = 0.685$--$0.764$.
We attribute this to time-variability associated with both
$\phi$- and $\Psi$-phase, and possibly luminosity fluctuations.

\citet[]{uv_disk_lines} found that the UV flux peaked at orbital phase
$\phi = 0.5$, and interpreted this as the illuminated face of HZ~Her.
We can use this information to set a limit to the distance between
the pulsar and the narrow line region ($r_{\rm line}$).
Since $r_{\rm line} \lesssim \sqrt{ d_{\rm crit}^2 + r_{\rm L1}^2 }$,
and $d_{\rm crit} > r_{\rm L1} \sim 3 \times 10^{11}$~cm,
we estimate $r_{\rm line} \lesssim d_{\rm crit}$, where $r_{\rm L1}$
is the distance between Her X-1 and the L1 Lagrange point.
This limit on $r_{\rm line}$, obtained from
photoexcitation, is shown by the vertical line in Figure \ref{fig:loclim1}.

\subsection{Limits on the density from thermal and ionization balance}
\label{sub:param}

Because the $R$ ratio is insufficient to quantify the density, we
use the ionization parameter $\xi$ instead.
The ionization parameter is defined as
$\xi = L / n_e r^2$~erg~cm~s$^{-1}$ \cite[]{kmt}, where $L$ is the
luminosity, $n_e$ the electron density, and
$r$ is the distance from the neutron star to the X-ray line
emitting region. 
The ionization state of an optically thin, photoionization-dominated
gas in equilibrium is fully determined by $\xi$, and by the spectral shape of the ionizing
continuum.
Assuming thermal and ionization balance, and
with the main-on continuum as input,
we calculate the charge state distribution for a grid of $\xi$ 
by use of the XSTAR  plasma code \cite[]{kallman82}. 
Our XSTAR models and recombination emission calculations (section
\ref{sec:em}) determine the $\xi$ where the recombination flux from 
a given ion is significant. For a fixed emission measure (section \ref{sub:em}),
the $\xi$-values which correspond to the
FWHM of the recombination line power of \ion{Ne}{9}, \ion{O}{7}, and \ion{N}{6} 
He$\alpha$ lines are $\log_{10} \xi =$1.3--2.0, 1.0--1.9, and 0.9--1.7, respectively.

To partially account for the existence of optically thick gas in Her
X-1, we do not make density estimations with $\xi$ and instead set
limits which are more robust. We account for the decrease in number of
photons available for photoionization, for the case where 1) the line emission region is
optically thin, and 2) the line-of-sight from the line emission region to
the neutron star may be optically thick. We do not account for line
emission in an optically thick region, hence we do not need a true transfer computation.
One large source of uncertainty in the density estimates obtained from $\xi$ is 
our knowledge of the flux impinging on the line emission region.
In Her X-1, the X-ray flux variation with $\Psi$-phase is associated with large changes of the
optical depth in our line-of-sight to the pulsar.
Since the optical depth between the pulsar and the line emission region
can be $\tau > 0$, we set upper limits to the density from
$n_e = L e^{-\tau} / \xi r^2 < L_{\rm max} / \xi_{\rm min} r^2$, where $L_{\rm max}$ is
the main-on state, or maximum, X-ray luminosity.
We take $\xi_{\rm min} = 1.0$ for \ion{O}{7} from the T lower
limit set in section \ref{sub:rrc}, and from the
$\xi(T)$ solution from XSTAR. Because $\xi_{\rm min}$ coincides with
the lower end of the FWHM of the \ion{O}{7} He$\alpha$ 
recombination line power, our calculations are validated by the data.
We use $\xi_{\rm min}$ and $L_{\rm max}=3.8 \times 10^{37} (D/6.6~{\rm kpc})^2$~erg~s$^{-1}$ to obtain
the density upper limits for \ion{O}{7} shown in Figure \ref{fig:loclim1}.
The density limits for \ion{Ne}{9} and \ion{N}{6} are $\sim 0.50$ and $\sim 1.3$ times
those for \ion{O}{7}, respectively.
As the density decreases below the limit, $\tau$ increases
and the $\xi$ of the emission region increases.

\subsection{Radiative Recombination Continuum}
\label{sub:rrc}

Free electrons recombining with ions produce an RRC 
width proportional to $T$.
The observed RRC photons correspond to recombination to the K shell.
For a photoionized gas, the RRC will be narrow, since
$T \ll I$, where $I$ is the ionization energy \cite[]{liedahl}.

The \ion{N}{7} RRC at $\lambda = 18.59$~\AA \ 
is detected with a $> 10 ~ \sigma$ confidence, yielding
$(6.5 \pm 2.8) \times 10^{-5}$~photons~cm$^{-2}$~s$^{-1}$,
and $kT = 3.7 \pm 1.3$~eV (90\% confidence). 
We also detected the \ion{O}{7} RRC at $\lambda = 16.78$~\AA \ with a
$\sim 4 ~ \sigma$ confidence level in the combined short-on and low state
data. The \ion{O}{7} RRC is visible in the added spectrum of
the low and short-on states (Fig. \ref{fig:added_spec}).
The \ion{O}{7} RRC normalization is $(2.8^{+3.3}_{-2.5}) \times 
10^{-5}$~photons~cm$^{-2}$~s$^{-1}$, and $kT$ is 2--6~eV (this 90\% confidence interval
is used to set a density limit in section \ref{sub:param}).
The RRC temperatures confirm photoionization is the dominant heating
mechanism \cite[]{liedahl}. Other RRC are not detected.

The observed \ion{N}{7} RRC is not far from the expected flux,
if compared to \ion{N}{7} Ly$\alpha$. The \ion{N}{7} ion has the best
measured RRC-to-Ly$\alpha$ ratio, which is $0.28 \pm 0.14$. 
The hydrogen-like RRC should have $\sim 0.75$ the
flux \cite[]{liedahl} of the corresponding Ly$\alpha$ flux (see Table
\ref{tab:lines}). Thus, the \ion{N}{7} RRC is close to, but weaker,
than the expected strength. For both \ion{O}{7} and \ion{N}{7}, the
measured $T$ agree with our XSTAR calculations, within the statistical errors. 

A high-temperature component, or a finite optical depth, 
can explain the weak RRC observed in the narrow line region.
An RRC with $T \gtrsim 50$~eV, which may originate in an extended hot
corona (section \ref{sub:em}), would be difficult to distinguish from
the continuum.  Alternatively, if the optical depth is significant, RRC photons 
produce ionizations and "fill in" the absorption edges, 
thereby pumping the recombination lines. However, optical depth effects
also imply changes in the Ly$\alpha$/Ly$\beta$ ratios, and
perhaps on the He$\alpha$~$r$ and Ly$\alpha$ lines, which
are not observed.

\section{Spectroscopy of the main-on state}
\label{sec:mainon}

\begin{figure}
\plotone{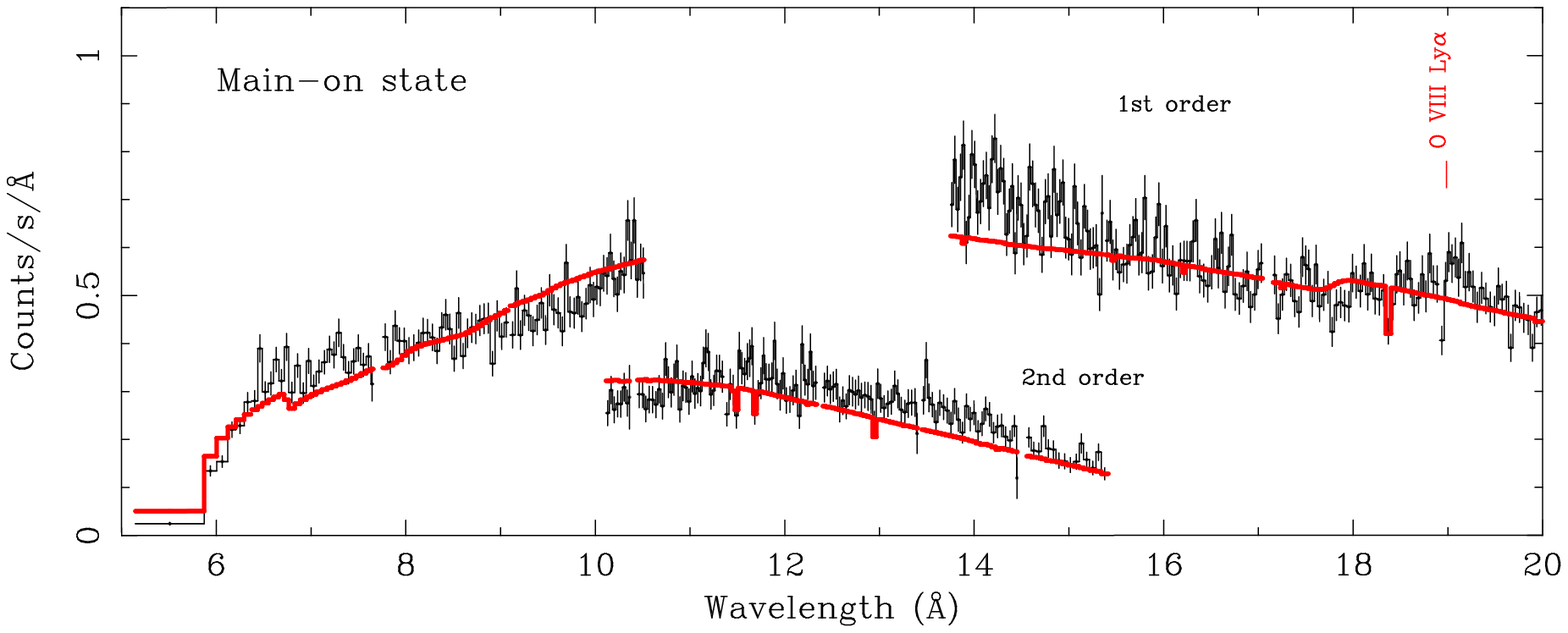}
\plotone{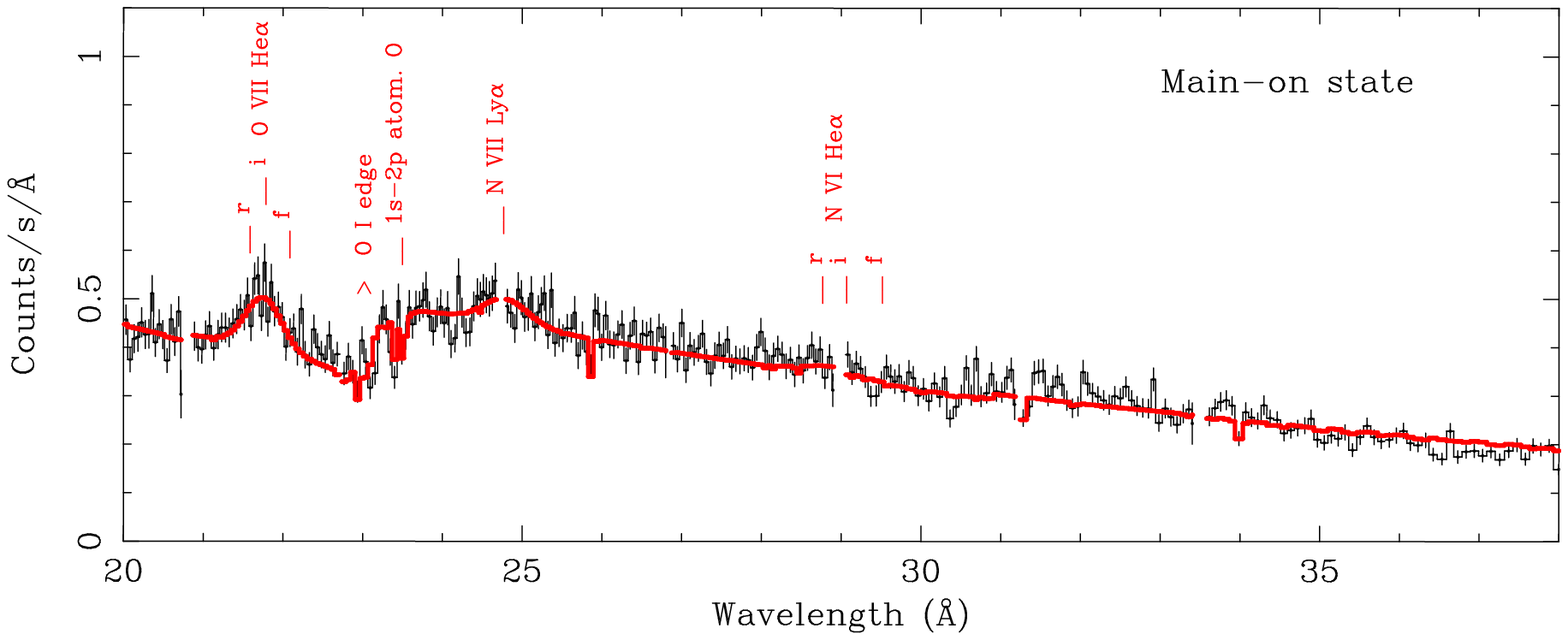}
\caption{High-resolution spectrum with RGS~1 for the main-on state observation
(11~ks), plus a phenomenological model fit (thick-line). Both first
and second order spectra are shown in the upper and lower part of the
figure, respectively.
\label{fig:high}}
\end{figure}

The main-on RGS spectrum is characterized by a bright continuum which has a
broad "bump" at $\sim$10--15~\AA \ (Fig. \ref{fig:high}). Excess
emission near the \ion{O}{7} He$\alpha$ wavelength is present.
We performed a single power-law fit on the spectrum, 
using the first order data, and we filled the wavelength gap with the second order data.

The RGS spectrum is dominated by a soft continuum component which is
distinct from the hard (photon index = 0.87) power-law component
observed with EPIC \cite[]{ramsay}. This hard power-law component 
does not significantly contribute to the RGS spectrum for $\lambda
\gtrsim 8$~\AA. For simplicity, we fit the RGS continuum level independently 
(Table \ref{tab:cont}).  The EPIC PN is consistent with the RGS spectrum
at $5 < \lambda < 14$~\AA, within the known uncertainties in the calibration.

\begin{figure}
\epsscale{0.7}
\plotone{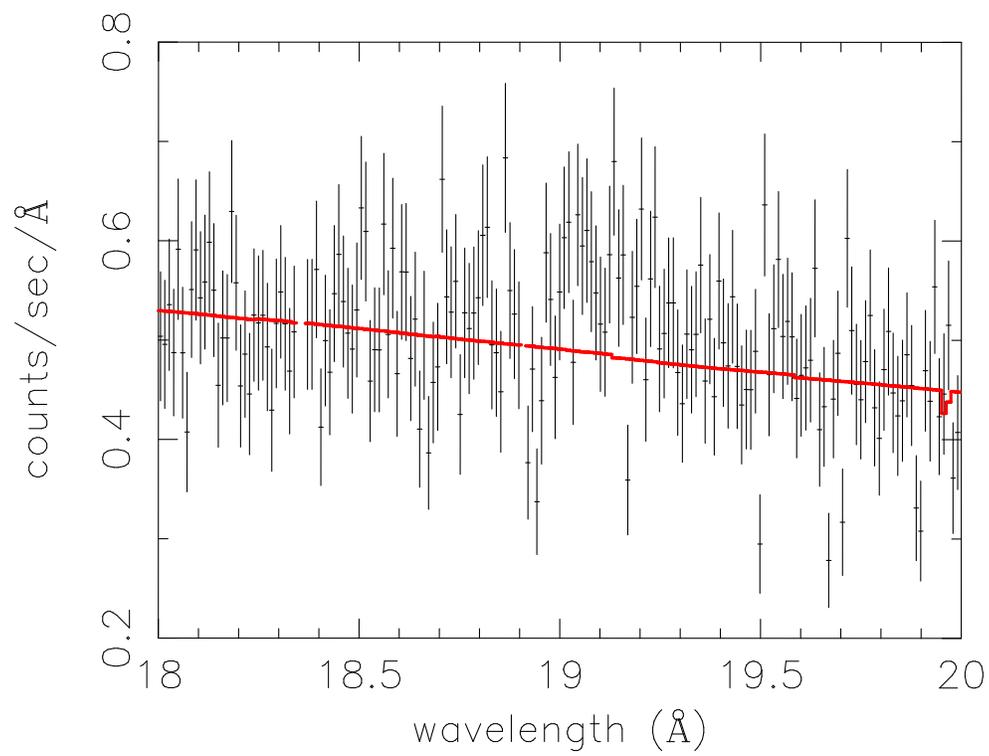}
\caption{Main-on observation near the \ion{O}{8} Ly$\alpha$ wavelength of 18.97~\AA, at
full resolution. Further observations are needed to determine whether
there is a P~Cygni feature.
\label{fig:pcyg}}
\end{figure}

Centered at the \ion{O}{7} $i$ line wavelength, we observe
an excess of emission (see Fig. \ref{fig:ovii}).  Due to the
noise in this feature, we are not certain if it is composed of
a single line or multiple lines, or if it is an emission line at all. 
However, rebinning reveals a statistically significant emission feature.
For reference, we perform a two-Gaussian line fit,
assuming the feature is \ion{O}{7} He$\alpha$, with $R=0$ and
$G=4$ (the line ratios observed in the other states). This yields 
a $\sigma = 3200 \pm 800$~km~s$^{-1}$ velocity broadening, but we caution
that $\sigma$ may not have a physical meaning. We search for a feature
at the \ion{O}{8} Ly$\alpha$ wavelength, 
but the statistics are not
sufficient for a detection (see the observed profile in Fig.
\ref{fig:pcyg}). We set upper limits for the \ion{O}{8} Ly$\alpha$ and
\ion{N}{7} Ly$\alpha$ line fluxes, assuming they have the same
velocity broadening as the putative \ion{O}{7} He$\alpha$.  

We detect a neutral O edge and set a limit on the neutral N edge (see
edge data in Table \ref{tab:oedge}). The RGS calibration data in the
SAS version 5.3 analysis software does include an instrumental O
(oxide) edge. Cold absorption models with solar abundances do not
provide acceptable fits. We fit the O edge to obtain an upper limit
on the interstellar \ion{O}{1} column density of $N_{\rm O~I } < 4.8 
\times 10^{17}$~cm$^{-2}$, which implies an equivalent H column of
$(N_{\rm H})_{\rm eq} < 5.7 \times 10^{20}$~cm$^{-2}$, if we use the
\cite{wilms} solar abundances. A resonant 1s-2p absorption feature
from atomic O is detected. This 1s-2p interstellar absorption feature 
has been observed before in other X-ray binary spectra \cite[]{paerels}.
We set a 90\% upper limit for the column $N_{\rm N~I } < 1.2 \times
10^{17}$~cm$^{-2}$.  The Ne and Mg edges are not detected, but limits
cannot be set reliably due to the uncertainty in the continuum in the
$\sim$10--15~\AA \ range.

Analysis of the first and second order RGS~1 spectrum does not
yield any robust line identifications of individual Fe L transitions.
A power-law fit results in negative residuals at 9--10~\AA, and positive residuals
at 11--15~\AA \ and $\sim 7$--8~\AA (Fig. \ref{fig:high}). 
These residuals do not appear thermal, and they
can be described as a broad "bump" at $\sim$10--15~\AA.
The "bump" at 10--15~\AA \ was previously identified with Fe L
emission \cite[]{mainon_lecs_beppo}, but these observations were performed
with low-resolution instruments.  The high-resolution spectrum
obtained by the {\it Einstein} spectrometer also showed a broad
emission feature at 1~keV with low statistical significance
\cite[]{grating_einstein,highres_spec_binaries_einstein}.  The
possible nature of this "bump" is discussed in section \ref{sub:bump}.

\section{A method to measure elemental abundance ratios 
from recombination emission lines} 
\label{sec:em}

The He$\alpha$ and Ly$\alpha$ emission lines observed in the
low and short-on states are modeled to extract the elemental abundance
ratios between C, N, O, and Ne. We calculate the 
\ion{C}{6} Ly$\alpha$, \ion{N}{6} He$\alpha$, \ion{N}{7} Ly$\alpha$,
\ion{O}{7} He$\alpha$, \ion{O}{8} Ly$\alpha$, and \ion{Ne}{9} He$\alpha$
line fluxes from a grid of
regions which have $\xi$ separated by $\Delta(\log_{10} \xi) = 0.1$. Good
$\xi$-sampling is necessary for an accurate abundance determination because
the recombination rates and charge state distribution are sensitive to $\xi$. 
To calculate the state of the gas in each region, we run
the XSTAR code, using as input the main-on broadband continuum shape.
% revised
The main-on continuum consists of a 92~eV blackbody, a broken
power law which has a break at 18 keV and photon indeces 0.88 and 1.8, 
and a 24 keV exponential cutoff.
This spectrum was measured by \citet[]{beppo_cyc},
and it is consistent with the EPIC observations.
We assume thermal and ionization balance, which is justified due to
the short time required for equilibration at the densities of interest
(see the limits on the
density in section \ref{sub:param}).  Knowledge of the emission
measure $EM = \int n_e^2 dV$ as a function of $\xi$ is necessary to
measure the elemental abundances, which implies a model dependence.
%revised
To constrain $EM(\xi)$ from the spectrum, one may choose a single $\xi$ for
each ion, such that the radiative recombination flux is maximized, and then
calculate the $EM$ for all the ions detected \cite[]{sako}.
However, each ion species emits copiously at a
broad range of $\xi$, which must be taken into account to obtain an
accurate measure of $EM(\xi)$ and the abundances.

We introduce a method to find the 
emission measure and the elemental
abundances simultaneously.
We assume a functional form for $EM(\xi)$ and then 
test the likelihood of the hypothesis with $\chi^2$.
The differential emission measure distribution
is well fit with the power-law form 
\begin{equation}
\label{eq:dem}
DEM(\xi) = \frac{\partial(EM)}{\partial(\log_{10} \xi)} = K \xi^\gamma . 
\end{equation}
The zero-slope $\partial(EM) / \partial (\log_{10} \xi) = 0$,
and the commonly assumed
$\partial (EM) / \partial \xi = K ~ \delta(\xi - \xi_o)$
(this is equivalent to a single ionization parameter),
are inconsistent with the observed spectrum,
where $K$, $\gamma$, and $\xi_o$ are fit parameters and $\delta()$ is
the delta function.
We use the recombination
rate coefficients calculated 
by Liedahl (private communication)
with the $HULLAC$ code \cite[]{hullac}.
The observed fluxes are 
fitted using a simplex $\chi^2$-minimization algorithm,
with 5 free parameters: the relative abundances, $K$, and $\gamma$. 
We fit each observation independently,
to test the robustness of the method. 
The results in Table \ref{tab:elem} show that 1)
consistent fits can be found
by the algorithm, 2) the phenomenological 
equation (\ref{eq:dem}) provides an excellent
fit to the RGS spectra in the low state and a marginal fit in
the short-on state, 
3) the results are sensitive to the shape of the ionizing spectrum, which
can be seen by comparing the results obtained with the main-on
spectrum and with a 20~keV bremsstrahlung spectrum,
and 4) the N/O abundance ratio is at least four times the solar value,
while the C/O and Ne/O ratios indicate that both carbon and oxygen 
are depleted with respect to neon. The marginal fit of the short-on
spectrum may be due to a physical emission measure which differs from
a power-law distribution.  The large (\ion{N}{7} +
\ion{N}{6})/(\ion{O}{8} + \ion{O}{7}) ratio is evident in the spectrum
on Figure \ref{fig:data_mod_lowshort}.

We discuss the validity and limitations of the method.
The method is robust insofar the $DEM(\xi)$ is well-covered in $\xi$ by
the observed ions, and because
the emission measure is not extrapolated to other $\xi$.
Also, it relies on the observed absence of optical depth 
or resonance scattering effects in the helium-like ion line ratios.
The abundance ratios depend on the XSTAR plasma code assumptions,
which may introduce systematic errors that are not addressed in this paper.
We have used a 20~keV bremsstrahlung as a benchmark to test the
dependence of our results to the ionizing spectrum,
and we do not observe an improvement in the 
$\chi^2$ of the fit; albeit the C/O and N/O abundance ratios 
are lower than with the main-on spectrum.
XSTAR is run at $n_e = 10^{10}$ cm$^{-3}$, a regime for which 
three-body recombination of the observed ions is unimportant, and in the
optically thin case.
For $n_e < n_{\rm crit}$, the
recombination line fluxes depend only weakly on $n_e$.
The value of $n_{\rm crit}$ depends on the ion,
and is roughly $n_{\rm crit} \sim 10^{17}$ cm$^{-3}$ for \ion{O}{8} \cite[]{kallmanba}.
The low-density assumption holds for the narrow line region (Fig. \ref{fig:loclim1}).
We assume Gaussian statistics since there are $> 44$ photons in each
of the lines used for the abundance ratio measurement.  We excluded
\ion{Ne}{10} Ly$\alpha$ from the fit since it is detected with low
statistical significance, but our fits are consistent with the
observed limits. 

Models of the Her X-1 UV spectrum roughly agree with the abundances we measure.
CNO-processing abundances were also evidenced by the
\ion{N}{5}/\ion{O}{6} line ratio observed in the
UV \cite[]{hut_uvlines}, which was compared to spectral models
of a photoionized accretion disk \cite[]{raymond}.
We crudely estimate an abundance ratio N/O $\sim 6.0$ times solar from the
observed discrepancy of the \ion{N}{5}/\ion{O}{6} line flux ratio with the CNO-disk model.
However, the \ion{N}{5}/\ion{O}{5} ratio yields 
N/O $\sim 2.6$ times solar. A \ion{C}{4}/\ion{O}{5} ratio
depleted below the solar abundance value was also observed by \citet{hut_uvlines}.
Our X-ray spectral measurements are more reliable than the UV ones
because they are less sensitive to radiation transfer effects, and because the
X-ray line fluxes have been fitted consistently with the $DEM$.

The functional form of $DEM(\xi)$, through the $K$ and $\gamma$ parameters,
contains information on the gas density and geometry.
A power-law dependence of the $DEM(\xi)$ is expected in some simple
geometries such as: a uniformly filled sphere, an optically-thin disk atmosphere with
a scale height $H \propto r^n$, and an 
illuminated slab with $\rho(z) \propto \rho_o e^{-z/z_o}$.
A full interpretation of the $K$ and $\gamma$ parameters is
outside the scope of this paper. 
\newpage 

\section{Discussion}
A new, X-ray narrow line component in the Her X-1 system has been
discovered.  We discuss the possibility that this narrow line region
is the illuminated face of HZ~Her, or an accretion disk atmosphere and
corona. Our elemental abundance measurements serve to quantify the CNO
processing products from HZ Her.  We discuss the connection between
the CNO products and the history of the binary system.

The main-on spectrum shows a peculiar continuum feature and possibly
a broad emission line component. We discuss the physical plausibility
of such a broad line component.

\subsection{Recombination emission from the disk atmosphere or 
the illuminated companion}
\label{sub:em}

We investigate the nature of the photoionized gas emitting narrow X-ray
lines during the low and short-on states.  The UV line emission
consists on a narrow ($\sigma \sim 80$~km~s$^{-1}$)
line component, which is likely to originate on the illuminated
face of HZ~Her, plus a broader ($\sigma \sim 300$~km~s$^{-1}$)
component, attributable to the accretion disk \cite[]{uv_disk_lines}.
Similar dynamical components in the X-ray narrow line region
would not be distinguished with the RGS energy resolution alone. 

The limits found on the density and location of the narrow line region
are shown in Figure \ref{fig:loclim1}.  We set the density lower limit of the
\ion{O}{7} region with the emission measure calculated from the line flux.
The lower limit to the root-mean-square of the density is
$(n_{\rm rms})_{\rm min} = \sqrt{EM/V_{\rm max}}$, where
the maximum volume $V_{\rm max} = \frac{4}{3} \pi r^3$. 
The $R$-ratio density thresholds listed in section \ref{sub:helike}
indicate that, within much of the allowed region
in Figure \ref{fig:loclim1}, the $2^3S \to 2^3P$ electron-impact excitation rate
is larger than the photoexcitation rate.
Radius limits for the putative Kepler orbits are obtained from the circularization
radius and from upper limits to the velocity broadening (Fig. \ref{fig:loclim1}).
%revised
However, a disk larger than the circularization radius ($\sim 1.7
\times 10^{11}$~cm) was deduced from UV and optical light curves
\cite[]{howa}.

\begin{figure}
\epsscale{1}
\plotone{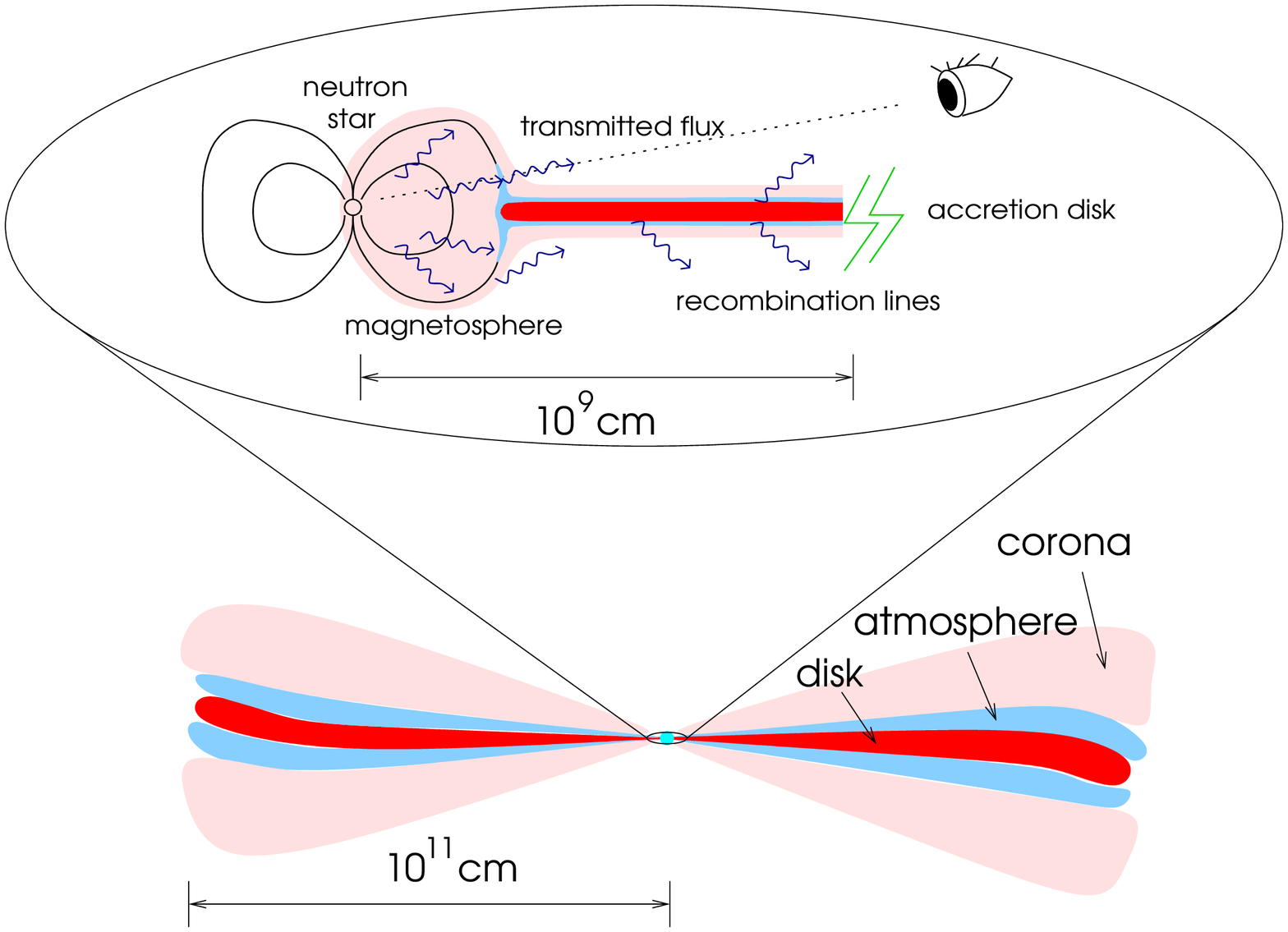}
\caption{Hypothetical sketch of an accreting, magnetized neutron star,
loosely based on the \cite{lamb} model. The 
accretion disk structure is also sketched, based
on models by \cite{jimenez}. We define the atmosphere as a thin skin
on the disk which is emitting X-ray recombination emission.  The disk
atmosphere and corona, or the illuminated face of the companion (not
shown), may be responsible for the narrow line emission. 
Plasma from the inner disk can be channeled by the magnetic field,
forming a highly ionized and possibly Compton-thick structure in
the magnetosphere. This structure and the inner disk are potential
sites for Compton scattering, X-ray absorption, and recombination
emission.  The tilt of the pulsar magnetic moment with respect to its
spin axis is not shown.
\label{fig:mag}}
\end{figure}

An extended, photoionized accretion disk atmosphere and corona may be
responsible for the observed recombination emission and the underlying
continuum. Model calculations \cite[]{jimenez} show that a centrally
illuminated, photoionized accretion disk would develop an extended,
Compton-temperature coronal structure, and a more compact, X-ray
recombination-emitting atmospheric layer (see Fig. \ref{fig:mag}).
We first interpret the behavior of the Her X-1 spectrum in
terms of this picture. The line fluxes observed are within a factor of
two (scaled by the Her X-1 luminosity and disk radius) of the
photoionized accretion disk atmosphere model results, and the line
ratios are all within a factor of unity of model values. The model
density of the helium-like ion region is $10^{13}$--$10^{14}$
cm$^{-3}$ if the disk radius is $r \sim 5 \times 10^{10}$ cm, which
are inside the limits on Figure \ref{fig:loclim1}. 
If the X-ray line velocity broadening can be resolved, this would support
the accretion disk emission hypothesis (the evidence is marginal from
the RGS, see Table \ref{tab:lines}).
The line emitting disk atmosphere is optically thin
towards most lines of sight, although its flattened geometry allows it to partially
shield itself from the pulsar ($e^{-\tau} \lesssim 0.1$).
The soft X-ray continuum is dominated by Compton scattering in the
disk corona during the low and short-on states. The optical depth of both
atmosphere and corona
decreases as the 1) radius increases, and as 2) the
height above the disk mid-plane increases.
At high inclination, the interior of the disk
is obscured by the outer disk edge. 
As the disk precesses out of its edge-on configuration, parts of the
disk which are close to the neutron star are less obstructed,
and deeper layers of the disk come into view, increasing both the atmospheric
line flux and the Compton scattered light from the corona.
This may explain the spectral evolution observed from the low to the
short-on state.  As the disk inclination decreases further in the
main-on state, the pulsar comes into view and overwhelms the narrow
line flux from the disk.  The duration of X-ray eclipse ingress or
egress at the low and short-on states is measured in the $1500$~s to
$3$~hr range \cite[]{eclipses,low_state_ecl_asca},
corresponding to source radii of $4 \times 10^{10}$~cm to $3 \times 10^{11}$~cm.
Eclipses of the X-ray lines are very likely to occur, since 
similar radii estimates have been obtained from the 
eclipse of the broad UV line emission region.
About $ \sim 0.1 \times 10^{-13}$~erg~s$^{-1}$~cm$^{-2}$~\AA$^{-1}$ or 
$\sim 6 $~\% of the UV continuum, together with the broad \ion{C}{4}
and \ion{N}{5} emission lines, are gradually eclipsed, indicating a
$\sim 10^{11}$~cm UV emission region \cite[]{uv_disk_lines}. 
Eclipse measurements, together with the X-ray spectrum and its variability,
and the photoionized disk models, lend support to the disk atmosphere
and corona origin of the X-ray emission during the low and short-on states. 

The spectrum of the accretion disk corona (ADC) source 4U~1822-371
\cite[]{4u1822} is very similar to the Her X-1 low and short-on
state spectra, confirming that ADC sources provide an edge-on view of the disk and
supporting the idea that the accretion disk in Her X-1 is precessing.

\begin{figure}
\epsscale{1}
\plotone{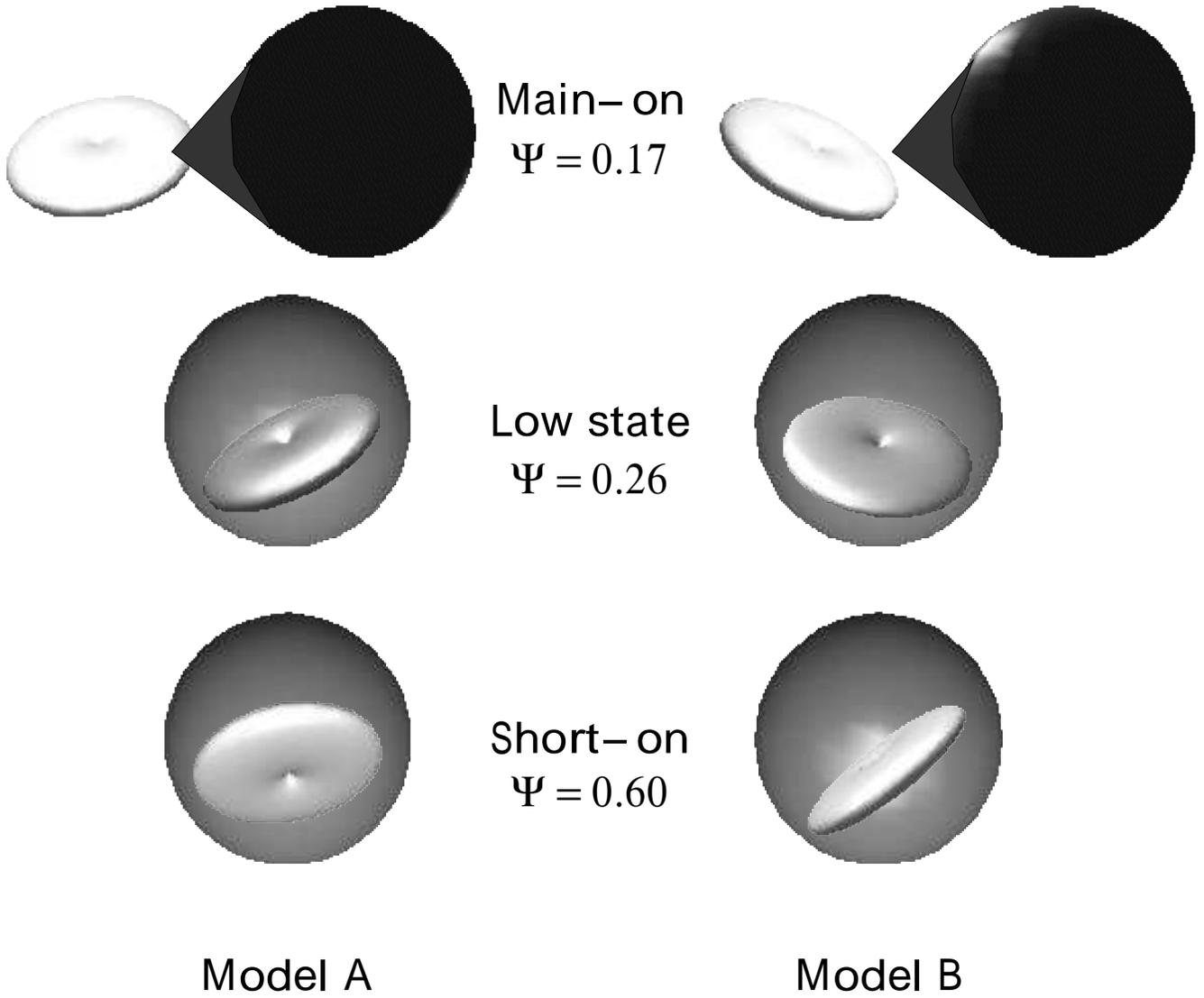}
\caption{Schematic of the Her X-1 geometry during our three
observations, showing the outer disk precession angle according to
Model A \cite[]{optical_curves,howa}, and to Model B \cite[]{disk_model}. 
The diagram does not include disk warp, is roughly to scale, and is
meant to show the projection of the system to our line of sight.
\label{fig:hergeo}}
\end{figure}

The variability of the X-ray line emission with $\Psi$-phase and $\phi$-phase, 
combined with the predicted disk inclination from the \cite{disk_model}
model, suggests that at least some of the X-ray line flux
originates on HZ~Her. The total line emission flux
in the RGS band at phase $\Psi=0.60$ is two times larger than
during phase $\Psi=0.26$, at approximately the same orbital phase
($\phi\sim 0.5$).
Taking the predicted outer disk inclination from the
\citet{disk_model} model, we have drawn the schematic in Figure
\ref{fig:hergeo} (Model B).
If the line emission originates in an accretion disk atmosphere and
corona, the line flux should increase as more of the disk's area comes
into view. If the line emission originates in HZ~Her,
the line flux should increase as the 
visible area of HZ~Her increases.
If the outer disk inclination predicted by the model of
\citet{disk_model} is correct, then some
of the line emission in the short-on must be coming
from the photoionized face of HZ~Her.
If the X-ray line emission originates on HZ~Her, it
should vary with orbital phase in a way similar to the UV continuum
flux \cite[]{uv_disk_lines}.
Some dependence of the narrow line flux on $\phi$-phase
is hinted by the main-on observation. The upper limits
to the narrow lines which we derive for the main-on observation
at $\phi \sim 0.2$, are below the line fluxes in the low
and short-on states, which were observed
at $\phi \sim 0.5$. However, this may be due to a 
$\Psi$-phase dependence only.
Observations at various orbital phases and similar $\Psi$-phase,
during the low and short-on states, can disentangle the
companion and disk components.

%revised
However, model values of the outer disk precession phase differ
(see Models A and B in Fig. \ref{fig:hergeo}).  Model B 
\cite[]{disk_model} yields a
precession phase which lags by $\Delta \Psi \sim 0.16$ 
the phase from model A \cite[]{optical_curves,howa},
which fits the optical and UV lightcurves. In model A,
the maximum disk opening angle to our line-of-sight occurs during the
main-on and short-on
peaks, at $\Psi =0.13$ and 0.63. Model A favors the
interpretation that the line emission originates in a disk atmosphere
and corona, since the projected disk opening angle and the line fluxes
during the short-on are larger than during the low state.

\subsection{New constraints on the X-ray binary evolution from the derived CNO abundances}
\label{sub:evol}

The unusually high N/O abundance ratio and the low C/O and O/Ne ratios
we have measured are the result of H-burning by the CNO-cycle in the
core of a massive star. The CNO-cycle produces N and depletes O and C,
until H is exhausted \cite[and references therein]{clayton}.  Evidence for
CNO processing in the Her X-1 system was previously obtained from UV
spectra, which show N-enrichment and C-depletion \cite[see section
\ref{sec:em}]{hut_uvlines}. Evolution models of an isolated star with
a mass comparable to HZ~Her cannot explain the observed abundances,
even after the end of the H-burning phase. A mechanism to enrich
the HZ~Her envelope with CNO-processed material is required, and such
a mechanism is likely linked to the presence of a companion. The CNO
abundances thus provide a stringent constraint on the evolution of the
binary system.

We compare the measured CNO yields with those in an isolated
intermediate-mass star.  Stellar evolution models \cite[]{stellar} of
an isolated $2.5~M_\sun$ star show a N/O abundance ratio enhancement
of 2.6 to 2.8 times the zero-age N/O ratio, a C/O ratio of 0.60 to
0.70 times its zero-age value, and an unchanged Ne/O ratio. The quoted
range corresponds to zero-age solar ($Z=0.02$) and sub-solar
($Z=0.001$) metallicities, at the end of the H-burning phase, after
$\sim 5.8 \times 10^8$~yr. The N-yield and the C-depletion are
maximized near the end of the H-burning phase. The
N-yield increases with stellar mass.  For example, a $4.0~M_\sun$ star
produces a N/O ratio which is 4.0 times its zero-age value
\cite[]{stellar}. Models of the evolution of an isolated $\sim
2.5~M_\sun$ star under-predict the observed N-yield in HZ~Her.

The observed N-enrichment with RGS cannot be attributed to systematic
uncertainties in the data, nor could we attribute it to
a dependence on the photoionization equilibrium calculations.
The systematic errors due to fitting of the emission lines and the
continuum are $< 25$\% (section \ref{sec:lowshort}).  Any RGS
calibration errors are $\sim 10$~\% where most of the emission lines
have been measured, and even the worst-case \ion{O}{1} edge
calibration error scenario implies a $\lesssim 35$\% error in the
\ion{O}{7} He$\alpha$ line flux, since $\tau_{\rm O~
I} \sim 0.3$. If we propagate a $\pm 35$\% uncertainty in the \ion{O}{7} He$\alpha$ flux through
the abundance extraction calculation, the N/O ratio changes by $< 35$\%.
The measured abundance ratios depend on the shape
of the ionizing spectrum, which affects the equilibrium state of the
plasma (section \ref{sec:em}). The changes in the 0.5-10~keV continuum
from the main-on state to the low and short-on states are well-fit with a varying
$N_H$ on a constant-slope power-law \cite[]{ramsay}.  Most of the line
emission should be generated by reprocessing of the direct 
pulsar (main-on) continuum, which consists of a power-law with a cutoff.
Taking a 20~keV bremsstrahlung ionizing spectrum still yields a
N/O~$>4.0$ times the solar value, and underabundant C and O with respect
to Ne (Table \ref{tab:elem}), so a systematic difference in the
plasma equilibrium state cannot, by this measure, explain the CNO
abundances.

The detection of CNO-processed material is validated by measurements
of the $({\rm C}+{\rm N}+{\rm O})/{\rm Ne}$ abundance ratio.  The 
$({\rm C}+{\rm N}+{\rm O})$ abundance sum is
conserved by the CNO-cycle, so the sum should be near the solar value.
We measure the $({\rm C}+{\rm N}+{\rm O})/{\rm Ne}$ 
ratio for the two best fits in Table
\ref{tab:elem}, obtaining $0.74 \pm 0.28$ times solar with the main-on
continuum model, and $0.45 \pm 0.19$ times solar with the 20
keV~bremsstrahlung model. 

To explain the observed abundances in HZ~Her, a mechanism is required to
transfer CNO-processed material to the envelope. One possibility is
that enhanced mixing and significant mass loss from the HZ~Her
envelope enrich it with CNO products from the core.
The stellar core is richer in CNO
products: for the $2.5~M_\sun$ example above, the abundance
ratios are C/O~$\sim 0.04$, N/O~$\sim 10$, and Ne/O~$\sim 10$ times the
zero-age values \cite[]{stellar}. These abundances have not reached
the equilibrium CNO values, which have higher N-yields.
The CNO products must have been efficiently transported
from the core to the surface.  However, the stellar core contains only 
$\sim 0.1$~times the mass of the star.  The shedding of a N-poor stellar envelope
though Roche-lobe overflow may explain the
observed N-enrichment. Intermediate-mass companions may survive
super-Eddington accretion, as evidenced by observations of the Cygnus
X-2 mass donor \cite[]{cygx2}, and by models
of the evolutionary track of that binary system \cite[]{cygx2mod}.
The mixing of the core's CNO products with the envelope gas 
is needed to explain the observations, but that may complicate the
calculation of the CNO burning rates, which depend on the
concentration of nuclear species in the core. 
Another possibility is that the CNO-processed material was transferred
from the neutron star progenitor onto HZ~Her. A massive neutron
star progenitor can readily produce the CNO elements in a $\sim 10^7$~yr
timescale, and would not require HZ~Her to be highly evolved.
The presence of a companion has likely affected the elemental
composition of HZ~Her, which presents a challenge to our understanding
of the evolution of the binary.

It is highly likely that the evolutionary state of the companion stars
in low- and intermediate-mass X-ray binaries is more evolved than had
previously been anticipated.  This has been shown by modeling the
evolution paths of 100 sample systems, from a grid of initial
stellar masses and periods \cite[]{rappaport}. 
These models show that many LMXB evolve from 
intermediate-mass binaries, so that many LMXB 
should be H-deficient and He-enriched, and 
the surface composition of the evolved secondaries should show
evidence of CNO processing.  The CNO abundances we find in Her X-1
are consistent with this basic picture.
The determination of the age of HZ~Her, its zero-age mass, and the
initial period of the binary can be better addressed by binary
evolution models now that the CNO abundances have been measured \cite[and private
communication]{rappaport}.

%%, which explains the unusually large Ne/O abundance ratio in many LMXB abs. spec.
\subsection{The physical plausibility of a broad line component}
\label{sub:magn}

We make the case that a broad line region is physically plausible
if it is neighboring the Her X-1 magnetopause, and we point out
that broad recombination lines have been observed previously
in another accreting X-ray pulsar.

A low-mass, compact X-ray binary system, 4U~1626-673, contains a
pulsar which is accreting from a white dwarf, and it exhibits
prominent recombination emission lines \cite[]{4u1626}.
The {\it Chandra} High-Energy Transmission Grating detected 
the \ion{Ne}{10} Ly$\alpha$, \ion{Ne}{9} He$\alpha$, \ion{O}{8}
Ly$\alpha$, and \ion{O}{7} He$\alpha$ lines
with FWHM~$\sim 2500$~km~s$^{-1}$, and it resolved them into
two distinct peaks, suggesting a disk origin. If the broadening
is due to the orbital velocity, the material would be located on the
disk, exterior to the pulsar magnetopause \cite[]{4u1626}. 
4U~1626-673 has a cyclotron line with $0.85$ times the energy than the 
corresponding line in Her X-1, a $3.4$ times larger corotation radius,
and broad lines with $\sim 1/20$th the flux of Her X-1 (the unabsorbed
X-ray flux from 4U~1626-673 is $\sim 1/36$th the
Her X-1 main-on flux). 
It is possible for broad recombination lines
to be associated with the accretion flows around X-ray pulsars, which
have $B \sim 10^{11}$--$10^{12}$~G. 

In Her X-1, from the putative \ion{O}{7} He$\alpha$ broad lines, we estimate that
the emission region would be inside the $4.5 \times 10^{8}~{\rm cm} < r < 3
\times 10^9$~cm boundary.  We assume the velocity broadening is due to
circular Kepler orbits, and that $\sim$90\% of the emission in the
\ion{O}{7} He$\alpha$ Gaussian line fit is enclosed within $v = \pm
2\sigma$ of the centroid. The estimated inner disk radius is just
larger than the corotation radius $r_{\rm co} = (G M P^{2}_{\rm pulse}
/ 4 \pi^2)^{1/3} = 1.9 \times 10^8$~cm, and is near the Alfv\`en radius \cite[]{lamb}.

The putative \ion{O}{7} He$\alpha$ broad line luminosity would imply 
that a fraction of order $\sim 10^{-1}$ of the luminous energy of the pulsar
is directed towards the line emission region. Emission measure calculations
allow us to set a $n_e > 10^{14}$~cm$^{-3}$ density lower limit,
from the radius upper limit above. This implies that 
there would be significant line opacity effects, which would have to be included in
the photoionization equilibrium calculations.
Hypothetically, a nearly Compton-thick region enshrouding the pulsar,
of radius of a few times $10^8$~cm, may be illuminating an inner
accretion disk, which then emits the broad recombination lines (Fig.
\ref{fig:mag}). To illustrate, a shell of Compton-thick gas at $r \sim
10^8$~cm, and with $\Delta r \sim 10^7$~cm thickness requires $n_e
\sim 10^{17}$~cm$^{-3}$ and $\log_{10} \xi \sim 4$, which indicates
the plasma would be fully ionized, except perhaps for iron.
An illuminated disk region of radius $r \sim 3 \times
10^9$~cm would have $n_e \sim 4 \times 10^{17}$~cm$^{-3}$, which
corresponds to the $\xi$ for maximum line power, and taking
$\Delta r \simeq 0.5 r$ and volume $V = 2 \pi r h \Delta
r$, its height would be $h \sim 10^3$~cm.

The duration of eclipse ingress and egress varies during the Her X-1 main-on state,
and it yields source size estimates in the $5 \times 10^8$~cm to $3 \times 10^9$~cm range,
albeit the smaller estimate may be affected by the HZ~Her atmosphere 
\cite[]{eclipses,eclipses_hz,eclipses_day}. 

%\cite[]{eclipses,eclipses_day,ginga_eclipse_circ_matter,low_state_ecl_asca},
 
\subsection{The 10--15 \AA \ "bump"}
\label{sub:bump}

The origin of the 10--15 \AA \ "bump" observed in the main-on state is
unclear. We describe three possible explanations for this "bump".
The first possibility is that the
"bump" is a superposition of Fe L emission lines with a
velocity broadening as in \ion{O}{7} He$\alpha$.
The blended transitions may range from \ion{Fe}{17} 
through \ion{Fe}{24}, at 10--17~\AA.
However, photoionized gases have weak Fe L emission
relative to hydrogen-like and helium-like ions of low- and mid-$Z$
elements \cite[]{liedahlme}. If the "bump" is Fe L emission, it could
be due to shock-heated gas in the magnetopause, although the 
$G$ line ratio from \ion{O}{7} He$\alpha$ is not consistent with collisional
heating.

The EPIC data suggest that the "bump" results from the overlap of 
the pulsar power-law continuum with a reprocessed, soft component. 
Shortward of $\lambda \sim 6$~\AA, for the hard power-law,
the pulse profile consists of sharply
peaked pulses, indicating a $\lesssim 10^8$~cm emission region.
Longward of $\lambda \sim 14$~\AA, the pulse profile is
nearly sinusoidal, indicating a $\sim 10^9$~cm emission region \cite[]{ramsay}.
The soft component is not a Planck spectrum.

A third possibility is that we are observing a blended and broad Fe L resonance
absorption complex. Fe L absorption is not suppressed in photoionized
gases. In this case, absorption lines from hydrogenic and helium-like
ions should also be observed.

\section{Conclusions}
\label{sec:concl}

We analyzed the high-resolution X-ray spectra
of Her X-1 obtained with the {\it XMM-Newton} RGS from
three observations performed at distinct states of its 35~d cycle.

We detect narrow recombination emission lines during the low and short-on states of
Her X-1.  Emission lines are detected from \ion{C}{6},
\ion{N}{6}, \ion{N}{7}, \ion{O}{7}, \ion{O}{8}, \ion{Ne}{9}, and
perhaps \ion{Ne}{10}, plus weak RRC of \ion{O}{7} and \ion{N}{7},
which indicate the gas is photoionized. The velocity broadening of the
lines is at the resolution limit of the RGS (e.g. $\sigma
\lesssim 260$~km~s$^{-1}$ for \ion{N}{7} He$\alpha$).
In the 5--38~\AA \ band, the low and short-on states have power-law
continua, while the main-on continuum exhibits excess
emission in the $\sim10$--15~\AA \ band.
The continuum flux during the short-on is twice as during the low state,
a tendency that is followed by most line fluxes as well.

We measure the abundance ratios among C, N, O, and Ne 
with the following method: we use a power-law emission measure
distribution, in conjunction with a grid of XSTAR plasma models,
and the $HULLAC$ recombination rates, to fit the narrow line fluxes, 
and obtain the abundance ratios with the emission measure parameters
simultaneously. We perform the measurements for the low and short-on
state data separately, and we obtain consistent results with
acceptable $\chi^2$ (Table \ref{tab:lines}). The enrichment
of nitrogen relative to oxygen, of more than four times the
solar values, plus the depletion of C and O with respect to
Ne, indicate extensive H-burning in a massive star.  The measured
abundances require a mechanism for transferring the CNO-processed
material onto the HZ~Her envelope. This mechanism is likely linked
to the presence of a companion, providing
an additional constraint on the evolution of the X-ray binary.

\begin{figure}
\epsscale{1}
\plotone{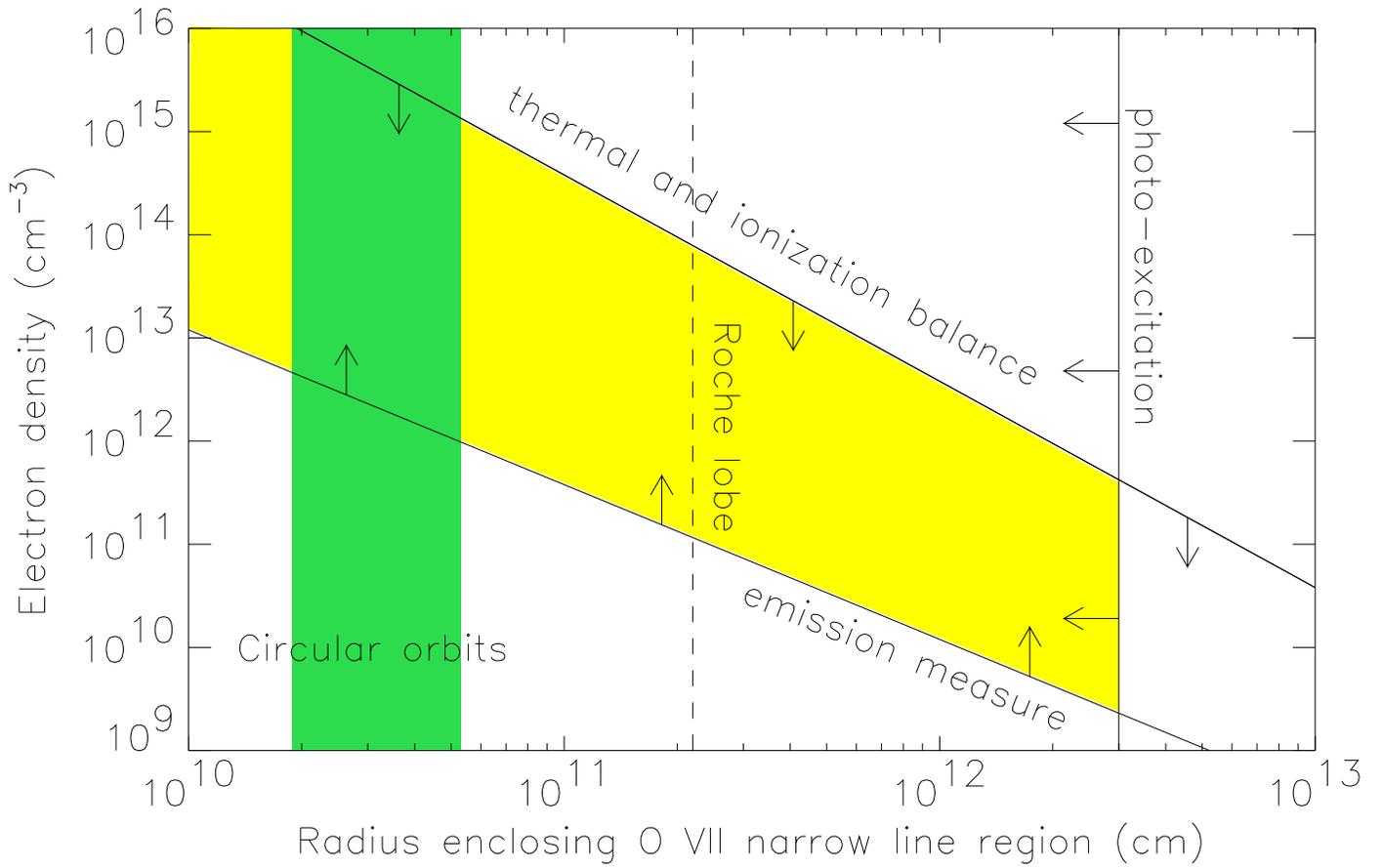}
\caption{Confidence limits (90\%) on the density and position of the narrow
line emission region, obtained from spectroscopy and modeling. The
limits shown correspond to \ion{O}{7}, and similar limits with
\ion{N}{6} and \ion{Ne}{9} can be set. 
\label{fig:loclim1}}
\end{figure}

We use spectroscopic analysis and models to set limits on the
density and location of the narrow 
line region. These limits (Fig. \ref{fig:loclim1})
provide important clues on the nature of
the line emission region. We assume
thermal and ionization balance to set upper limits on the
density, and we use the emission measure derived from the line fluxes to set
lower limits on the density. If the line velocity broadening 
is due to Kepler motion, we may set bounds to the orbital radii.
We use the He$\alpha$ line ratios and UV photoexcitation calculations
to set upper limits to the radius enclosing the region.  The low
RRC temperatures ($30,000 < T < 60,000$~K) allow us to validate
our density upper limits and the photoionization equilibrium
models.

The narrow line region may be identified with an accretion disk
atmosphere and corona, or with the illuminated face of HZ~Her.
The evidence for the disk identification relies on the modeled structure and spectra from
a photoionized disk \cite[]{jimenez}, which agree with the limits set
on the density ($10^{13}$--$10^{14}$~cm$^{-3}$).
The unresolved velocity broadening indicates the outermost 
($r \sim 5 \times 10^{10}$ cm) radii of the disk dominate the
emission, in agreement with the models, while fluxes of 
the observed hydrogen and helium-like lines are within a factor
of two of the calculations.  Observations from previous missions of
the eclipse ingress and egress reveal an emission region which can
match the size of the disk and its corona.  The flux variation with
orbital phase and with 35~d phase, on the other hand, favors a contribution from the
illuminated face of HZ~Her to the narrow line emission.

The variability of the Her X-1 spectrum lends support to the precession
of the accretion disk. Regions with contrasting dynamical properties
are coming into view at different 35~d phases. Notably, Her X-1
exhibits an ADC-like spectrum during low and short-on states,
indicating an edge-on disk, while during main-on states, the spectrum
is dominated by the continuum, due to a disk
inclination which exposes the pulsar to our
line of sight.

We detect an excess of emission centered on the \ion{O}{7} He$\alpha$ $i$ line
wavelength during the main-on state. If this feature is due to the 
\ion{O}{7} He$\alpha$ $i$ and $r$ lines,
their velocity broadening would be in the order of $10^3$~km~s$^{-1}$. 
Similarly broad, yet double-peaked lines, have to date
only been observed from the accreting pulsar in 4U~1626-67 \cite[]{4u1626}.
The observed emission emission feature is noisy, but if it is real, it
may be due to emission in the inner accretion disk, near the pulsar
magnetopause.

\acknowledgments

We are grateful for the input provided by Duane Liedahl. We thank
Saul Rappaport, John Raymond, and Maurice Leutenegger for engaging in 
fruitful discussions.  We thank the members of the RGS instrument team and the
MIT CXC and HETG groups for their support.  We acknowledge our use 
of the public data which was made promptly available by the \it
RXTE \rm all-sky monitor team.

\clearpage

%% No more than seven \figcaption commands are allowed per page,
%% so if you have more than seven captions, insert a \clearpage

%% There must be a \figcaption command for each legend. Key the text of the
%% legend and the optional \label in curly braces. If you wish, you may
%% include the name of the corresponding Figure file in square brackets.
%% The label is for identification purposes only. It will not insert the
%% Figures themselves into the document.
%% If you want to include your art in the paper, use \plotone.
%% Refer to the on-line documentation for details.

\clearpage

\clearpage

\clearpage

\clearpage 

\clearpage

\clearpage

\clearpage

\clearpage

\clearpage

%% Tables should be submitted one per page, so put a \clearpage before

%% Two options are available to the author for producing tables:  the
%% deluxetable environment provided by the AASTeX package or the LaTeX
%% table environment.  Use of deluxetable is preferred.
%%

%% Three table samples follow, two marked up in the deluxetable environment,
%% one marked up as a LaTeX table.

%% In this first example, note that the \tabletypesize{}
%% command has been used to reduce the font size of the table.
%% Note also that the \label command needs to be placed 

\clearpage

\begin{table}
\begin{center}
\scriptsize
\caption{Observations with the Reflection Grating Spectrometer \label{tab:obs}}
\begin{tabular}{cccccccc}
\\
\tableline
\tableline
 MJD\tablenotemark{a} & 
{\it XMM} & Observation & Exposure & Orbital  & 35 d & RGS 1 & RGS 2 \\
Interval & Orbit & Start & (ks) & Phase$^{(1)}$ & Phase\tablenotemark{b} & (c/s)\tablenotemark{c} & (c/s)\tablenotemark{c} \\
\tableline
 51935.036--51935.169 & 207 & 2001 Jan 26, 00:51:13~UT & 11.31 &  0.18--0.26 & 0.17 (main-on) & 11.29 & \nodata \\
 51972.927--51973.083 & 226 & 2001 Mar 4, 22:14:18~UT & 13.41 &  0.47--0.56 & 0.26 (low state) & 0.37 & 0.42 \\
 51984.906--51985.053 & 232 & 2001 Mar 16, 21:45:01~UT & 12.63 &  0.52--0.60 & 0.60 (short-on) & 0.83 & 0.93 \\
\tableline
\end{tabular}

%% Any table notes must follow the \end{tabular} command.

\tablenotetext{a}{Mean Julian Day.}
\tablenotetext{b}{The zero of the 35 d phase is obtained at MJD 51929.08 ($\phi_{\rm orb}=0.68$) and
MJD 51964.00 ($\phi_{\rm orb}=0.22$) from \it RXTE \rm $ASM$ dwell-by-dwell light curve data.
The $ASM$ light curve shows some evidence (a $4\sigma$ detection in one dwell)
that the second turn-on occurred at MJD 51963.08 ($\phi_{\rm orb}=0.68$)
instead, but is affected by pre-eclipse dips.}
\tablenotetext{c}{First order count rates.}
\tablerefs{(1) Deeter et al. 1998.}
\end{center}
\end{table}

\begin{deluxetable}{crrrrrrr}
\tabletypesize{\scriptsize}
\tablecaption{Observed X-ray emission lines
\label{tab:lines}}
\tablewidth{0pt}
\tablehead{
\colhead{Line(s)} & \colhead{$\lambda_{\rm o}$ (\AA)} 
& $\lambda$ (\AA) & $\sigma$ (km/s) & Flux\tablenotemark{a}
& \colhead{State}   &
\colhead{EW (eV)\tablenotemark{b}} & \colhead{Counts \tablenotemark{c}}
}
\startdata
C VI Ly$\alpha$ & 33.74       & $33.72 \pm 0.01$ & $< 380$     & $6.4^{+2.1}_{-1.8}$ &  Low & $ 2.7 \pm 0.4 $  &    47 \\
                &             & $33.75 \pm 0.02$ & $< 320$     & $10^{+4}_{-3}$  & Short-on & $ 1.8 \pm 0.3 $  &   66 \\ 
N VI He$\alpha$ & 29.53 ($f$) & \nodata              & \nodata            & $< 0.94$          &  Low & $< 0.6$        & \nodata \\
                       &      & \nodata              & \nodata & $< 1.6$      &  Short-on & $< 0.4$        & \nodata \\
                & 29.08 ($i$) & $29.07 \pm 0.01$ & $270 \pm 100$ & $29 \pm 4$          &  Low & $ 18.6 \pm 1.9$\tablenotemark{d} &  173\tablenotemark{d} \\
                       &      & $29.10 \pm 0.015$ & $< 260$     & $40 \pm 6.5$\tablenotemark{e} &  Short-on & $ 10.9 \pm 0.9$\tablenotemark{d}  & 217\tablenotemark{d} \\
                       &      & \nodata & \nodata & $\lesssim 30$ 	   &  Main-on & \nodata & \nodata \\
                & 28.78 ($r$) & $28.77 \pm 0.01$ & $< 260$     & $6.6 \pm 2.0$ &  Low & \nodata & \nodata \\
                       &      & $28.75 \pm 0.02$ & $< 260$     & $9 \pm 5$\tablenotemark{e} & Short-on & \nodata & \nodata \\
N VII Ly$\alpha$ & 24.78    & $24.77^{+0.02}_{-0.01}$ & $< 590$ & $15 \pm 3$   &         Low & $ 5.2 \pm 0.7 $  &  58 \\
                   &    & $24.80 \pm 0.02$ & $520^{+230}_{-180}$ & $23 \pm 5.5$\tablenotemark{e} & Short-on & $ 4.8 \pm 0.5 $  & 111 \\
                   &    & $24.85 \pm 0.17$ & 3200 (fixed) & $84 \pm 30$  & Main-on & \nodata & \nodata \\
O VII He$\alpha$ & 22.10 ($f$) & \nodata & \nodata & $< 3.6$         &  Low   & $< 2.8$        & \nodata \\
                      &        & \nodata & \nodata & $< 3.6$      &  Short-on & \nodata   &  \nodata \\
                 & 21.80 ($i$) & $21.79 \pm 0.01$ & $< 400$    & $31 \pm 5$          &  Low & $ 23.4 \pm 2.4 $\tablenotemark{d} & 154\tablenotemark{d} \\
                      &        & $21.82 \pm 0.02$ & $320^{+140}_{-160}$ & $70 \pm 10$ &  Short-on & $ 22.4 \pm 1.6 $\tablenotemark{d}   & 287\tablenotemark{d} \\
                      &        & \nodata & $3200 \pm 800 $ & $155 \pm 30$     &   Main-on & \nodata & \nodata \\
                 & 21.60 ($r$) & $21.61\pm 0.03$ & $< 1600$    & $10 \pm 4$ &  Low & \nodata & \nodata \\
                      &        & $21.62\pm 0.02$ & \nodata & $8 \pm 4 $   &  Short-on & \nodata & \nodata \\
                      &        & \nodata & $3200 \pm 800$ & $38 \pm 8$     &   Main-on & \nodata & \nodata \\
O VIII Ly$\alpha$& 18.97       & $18.96 \pm 0.01$ & $390 \pm 200$ & $12.0 \pm 2.5$ &  Low & $7.5 \pm 0.8 $   & 98 \\
                      &    & $18.96^{+0.02}_{-0.01}$ & $< 450$  & $14 \pm 3$    &   Short-on & $ 3.0 \pm 0.3 $  & 85 \\
                      &    & $19.05 \pm 0.05$ & P Cygni?  &  $\lesssim 90$   &   Main-on & \nodata & \nodata \\
Ne IX He$\alpha$ & 13.70 ($f$) & \nodata & \nodata & $< 1.6$            & Low & \nodata &  \nodata \\
                   &           & \nodata & \nodata & $< 6.2$ &  Short-on   & \nodata & \nodata \\
                 & 13.55 ($i$) & \nodata & \nodata & $5.6^{+2.5}_{-2.1}$  & Low & $ 10.1 \pm 1.7$\tablenotemark{d} & 44\tablenotemark{d} \\
                   &           & \nodata & \nodata & $12^{+3}_{-4}$ &  Short-on & $ 7.2 \pm 1.0 $\tablenotemark{d} & 61\tablenotemark{d} \\
                & 13.45 ($r$)  & \nodata & \nodata & $< 4.8$            & Low & \nodata & \nodata \\
                   &           & \nodata & \nodata & $3.9^{+3.1}_{-2.2}$ &  Short-on &  \nodata &  \nodata \\
Ne X Ly$\alpha$ & 12.13        & $12.20 \pm 0.04$  & \nodata & $7 \pm 3$  & Low & \nodata & \nodata \\
                &              & \nodata & \nodata & $3.6 \pm 3.0$  & Short-on  & \nodata & \nodata \\
 \enddata

%% Text for table notes should follow after the \enddata but before
%% the \end{deluxetable}. Make sure there is at least one \tablenotemark
%% in the table for each \tablenotetext.
\tablecomments{Errors are 90 \% confidence limits unless otherwise noted. 
Symbols: $\lambda_{\rm o}$ = theoretical wavelength;
$\lambda$ = measured wavelength; $\sigma$ = standard deviation;
State = Flux state within the Her X-1 35 d cycle; $EW$ = equivalent width.
Note $\lambda$, $\sigma$ and the flux are obtained from Gaussian fits.  }
\tablenotetext{a}{In units of 10$^{-5}$ phot cm$^{-2}$s$^{-1}$. The errors do not include the continuum level uncertainty, which
can be significant for the weakest lines.}
\tablenotetext{b}{Quoted errors in this column are 1-$\sigma$ (68 \% confidence limits).}
\tablenotetext{c}{After pipeline processing and application of
response matrices, these are the corrected X-ray counts per line (or line complex) with
the local continuum subtracted. The $EW$ is then calculated.}
\tablenotetext{d}{These measurements include both intercombination ($i$) and resonance ($r$) lines.}
\tablenotetext{e}{We use only RGS~2 data in these line fits, to avoid a few flagged pixels in RGS~1. The RGS~1 still detects these lines at levels comparable to RGS~2.}

\end{deluxetable}

\begin{table}
\begin{center}
\small
\caption{Continuum power-law fit parameters \label{tab:cont}}
\begin{tabular}{ccccc}
\\
\tableline\tableline
State 	& Photon Index  				& Normalization at 1~keV & $N_{H}$  & $\chi^2$/DOF \\
& &  (phot~keV$^{-1}$ cm$^{-2}$ s$^{-1}$)		& ($10^{18}$ cm$^{2}$)  & \\
\tableline
Low       & $1.65 \pm 0.06$ & $(4.1 \pm 0.1) \times 10^{-3}$ 	& $< 86 $				& $562/467$\tablenotemark{a} \ $=1.21$ \\
Short-on  & $1.75 \pm 0.03$ & $(9.8 \pm 0.2) \times 10^{-3}$				& $< 75$ 	& $1179/945$\tablenotemark{a} \ $= 1.25$ \\
Main-on   & $1.88 \pm 0.10$ & $0.130 \pm 0.007$ & $< 57$ 								& $4312/3122=1.39$ \\
\tableline
\end{tabular}

%% Any table notes must follow the \end{tabular} command.
\tablecomments{Errors are 90\% confidence limits. DOF = degrees of freedom.}
\tablenotetext{a}{The DOF reflect rebinning performed for plotting and fitting the continuum.}
\end{center}
\end{table}

%% If the table is more than one page long, the width of the table can vary
%% from page to page when the default \tablewidth is used, as below.  The
%% individual table widths for each page will be written to the log file; a
%% maximum tablewidth for the table can be computed from these values.
%% The \tablewidth argument can then be reset and the file reprocessed, so
%% that the table is of uniform width throughout. Try getting the widths
%% from the log file and changing the \tablewidth parameter to see how
%% adjusting this value affects table formatting.

%% In this example, we have used the optional * argument to \\ to
%% instruct LaTeX to keep rows together on the same page. (See the
%% lines following the \cutinhead.) Using \\* to group together table
%% rows on the same page affects how the table breaks. Try taking
%% the *'s out and LaTeXing again to see the difference.

\begin{table}
\begin{center}
\scriptsize
\caption{Helium-like line diagnostics and photoexcitation of the $2^3S$ level 
\label{tab:helike}} 
\begin{tabular}{ccccccccc}
\\
\tableline\tableline
    & Her X-1 & $R=f/i$  & $G=(f+i)/r$ & $\lambda_{\rm f \to i}$ & Flux & $w_{\rm f}$  & $w_{\rm f \to i}$($r$=$10^{11}$cm) & $d_{\rm crit}$ \\
Ion & State & Line   & Line  &  (\AA)$^{(1)}$ & $F_{\lambda_{\rm f \to i}}$ ($10^{-13}$ erg  & $2^3S \to 1^1S$ & $2^3S \to 2^3P$ & Radius \\
    &       & Ratio  & Ratio &  					 & cm$^{-2}$ s$^{-1}$ \AA$^{-1}$ )     & Rate ($s^{-1}$)$^{(2)}$ & Rate ($s^{-1}$) & (cm) \\
\tableline
Ne IX  & low		& $< 0.3$  & $> 1.3$       & 1270 & $1.9 \pm 0.1^{(3)}$ & $1.09 \times 10^4$ & $5 \times 10^5$ & $7 \times 10^{11}$ \\
       & short-on	& $< 0.5$  & $3.8^{+16}_{-2.1}$  &      &  \\
O VII  & low		& $< 0.14$  & $3.1 \pm 1.4$  & 1637 & $1.5 \pm 0.2^{(3)}$ & $1.04 \times 10^3$ & $1 \times 10^6$ & $3 \times 10^{12}$ \\
       & short-on	& $< 0.06$ & $8.7 \pm 4.5$  &      &    \\
%            & & &      &   & main-on \\
N VI   & low		& $< 0.04$ & $4.4 \pm 1.5$  & 1906 & $1.5 \pm 0.1^{(4)}$ & $2.53 \times 10^2$ & $2 \times 10^6$ & $9 \times 10^{12}$ \\
       & short-on	& $< 0.05$ & $4.4 \pm 2.6$  &      &   \\
\tableline
\end{tabular}

%% Any table notes must follow the \end{tabular} command.
\tablecomments{Errors are 90\% confidence limits. 
Symbols: $\lambda_{\rm f \to i}$ = wavelengths of $2^3S \to 2^3P$ transitions; $w_{\rm f \to i}$ = radiative decay rates;
$w_{\rm f \to i}$ = calculated photoexcitation rate.}
\tablerefs{(1) Porquet et al. 2001; (2) Drake 1971; (3) from the \it Hubble \rm $GHRS$ at $\phi=0.56$--$0.60$ 
by Boroson et al. 1996; (4) from the \it Hubble \rm $FOS$ at $\phi \sim 0.5$ by Anderson et al. 1994.}
\end{center}
\end{table}

\begin{table}
\begin{center}
\small
\caption{Absorption features \label{tab:oedge}}
\begin{tabular}{ccccc}
\\
\tableline\tableline
Edge 	 & $\lambda_{\rm o}$ (\AA)& $\lambda$ (\AA)  		& $\tau$ 		& State \\
\tableline
Neutral O  & 23.05       	 & $23.0 \pm 0.5 $ 			& $0.30 \pm 0.06$	    & Low \\ 
		 & 23.05       	 & $23.27^{+0.10}_{-0.13}$ 	& $0.28 \pm 0.04$ 	    & Short-on\\ 
		 & 23.05       	 & $23.1^{+0.1}_{-0.2}$  	& $0.22 \pm 0.02$ 	    & Main-on \\ 
Neutral N  & 30.77       	 &  (fixed)                  	& $< 0.08$	 	& Main-on \\ 
\tableline
Line 	 & $\lambda_{\rm o}$ (\AA)& $\lambda$ (\AA)  		& EW (eV) 		& State \\
\tableline
Atomic O 1s-2p  & $23.47^{(1)}$ & $23.50^{+0.03}_{-0.02}$ 	& $0.57 \pm 0.30$ 	& Main-on \\ 
\tableline
\tablecomments{Errors are 90\% confidence limits. Symbols: $\lambda_{\rm o}$ = theoretical
edge wavelength; $\lambda$ = measured wavelength; $\tau$ = edge optical depth; EW = equivalent width.}
\tablerefs{(1) McLaughlin \& Kirby 1998.}
\end{tabular}
\end{center}
\end{table}

%TABLE ON ELEMENT ABUNDANCES
\begin{table}
\begin{center}
\scriptsize
\caption{Elemental abundance measurements and differential emission measure ($DEM$) 
parameters 
\label{tab:elem}}
\begin{tabular}{lccccccc}
\\
\tableline\tableline
State 	& $\frac{({\rm C}/{\rm O})}{(C/O)_{\sun}}$ 	& $\frac{({\rm N}/{\rm O})}{({\rm N}/{\rm O})_{\sun}}$  & $\frac{({\rm Ne}/{\rm O})}{({\rm Ne}/{\rm O})_{\sun}}$  & $K$ 		& $\gamma$ & $\chi^2$ of Fit & Ionizing \\
		& 		&  		& 		& ($10^{58}$ cm$^{-3}$) & & (1 DOF) & Spectrum \\
\tableline
Low       & $0.52 \pm 0.12$ & $8.8 \pm 1.0$ & $1.9 \pm 0.7$ & $7.2 \pm 3.7$ 	& $1.59 \pm 0.17$ & 0.02 & P.L.+cut (main-on) \\
Short-on  & $0.56 \pm 0.15$ & $7.7 \pm 0.9$ & $2.5 \pm 0.7$ & $33^{+17}_{-11}$  & $2.05 \pm 0.19$ & 4.96 & P.L.+cut (main-on) \\
\tableline
Low       & $0.30 \pm 0.07$ & $5.9 \pm 0.7$ & $2.5 \pm 1.0$ & $10^{+4}_{-3}$ 	& $1.78 \pm 0.17$ & 1.25 & 20~keV brems. \\
Short-on  & $0.29 \pm 0.08$ & $4.6 \pm 0.6$ & $3.5 \pm 1.0$ & $43^{+17}_{-13}$  & $2.28 \pm 0.19$ & 8.72 & 20~keV brems. \\
\tableline
\end{tabular}
\tablecomments{We show the statistical errors at 66\% confidence. The element 
abundance ratios (by number of atoms) are normalized to the solar values compiled by
\citet[]{wilms}, which are (C/O)$_{\sun} = 0.47$, (N/O)$_{\sun} = 0.11$, and
(N${\rm e}$/O)$_{\sun} = 0.14$. The $K$ and $\gamma$ parameters define the $DEM$.}
\end{center}
\end{table}

%% If the table is more than one page long, the width of the table can vary
%% from page to page when the default \tablewidth is used, as below.  The
%% individual table widths for each page will be written to the log file; a
%% maximum tablewidth for the table can be computed from these values.

%\clearpage

%\begin{table}
%\begin{center}
%\small
%\caption{Emission measures.\label{emission measures}}
%\begin{tabular}{cccc}
%\tableline\tableline
%Ion & $R=f/i$\tablenotemark{a} & 
%$G=r/(f+i)$ & Density (cm$^{-3}$)\\
%\tableline
%\ion{Ne}{9} & & & \\
%\ion{O}{7} & & & \\
%\ion{N}{6} & & & \\
%\tableline
%\end{tabular}

%\tablenotetext{a}{Density diagnostic subject to photoexcitation degeneracy.}
%\end{center}
%\end{table}

\end{document}